\documentclass{article}
\usepackage{PRIMEarxiv}

\usepackage[utf8]{inputenc} 
\usepackage[T1]{fontenc}    
\usepackage{hyperref}       
\hypersetup{
    colorlinks=true,
    linkcolor=blue,
    filecolor=magenta,      
    urlcolor=cyan,
}

\usepackage{url}            
\usepackage{booktabs}       
\usepackage{amsfonts}       
\usepackage{amsmath}        
\usepackage{nicefrac}       
\usepackage{microtype}      
\usepackage{lipsum}
\usepackage{fancyhdr}       
\usepackage{graphicx}       
\usepackage{subfig}
\usepackage{soul}
\usepackage{lineno}
\usepackage{multirow}
\graphicspath{{media/}}     
\title{A Framework of Data Assimilation for Wind Flow Fields by Physics-informed Neural Networks
}

\author{
    Chang Yan\textsuperscript{1,2,3}\\
    \texttt{yanchang18@mails.ucas.ac.cn}\\
    \And
    Shengfeng Xu\textsuperscript{1,2}\\
    \texttt{xushengfeng21@mails.ucas.ac.cn}\\
    \And
    Zhenxu Sun\textsuperscript{1,*} \\
    \texttt{sunzhenxu@imech.ac.cn} \\
    \And
    Thorsten Lutz\textsuperscript{3} \\
    \texttt{lutz@iag.uni-stuttgart.de} \\
    \And
    Dilong Guo\textsuperscript{1,2} \\
    \texttt{guodl@imech.ac.cn} \\
    \And
    Guowei Yang\textsuperscript{1,2} \\
    \texttt{gwyang@imech.ac.cn} \\
}
\begin{document}
\maketitle
\vspace{-20pt}
\begin{center}
\textsuperscript{*}Corresponding author\\
\textsuperscript{1}Institute of Mechanics, Chinese Academy of Sciences, Beijing 100190, China \\
\textsuperscript{2}University of Chinese Academy of Sciences, Beijing 100049, China \\
\textsuperscript{3}University of Stuttgart, Institute of Aerodynamics and Gas Dynamics, 70569 Stuttgart, Germany
\end{center}
\vspace{20pt}

\begin{abstract}
Various types of measurement techniques, such as Light Detection and Ranging (LiDAR) devices, anemometers, and wind vanes, are extensively utilized in wind energy to characterize the inflow. 
However, these methods typically gather data at limited points within local wind fields, capturing only a fraction of the wind field's characteristics at wind turbine sites, thus hindering detailed wind field analysis.
This study introduces a framework using Physics-informed Neural Networks (PINNs) to assimilate diverse sensor data types.
This includes line-of-sight (LoS) wind speed, velocity magnitude and direction, velocity components, and pressure.
Moreover, the parameterized Navier-Stokes (N-S) equations are integrated as physical constraints, ensuring that the neural networks accurately represent atmospheric flow dynamics.
The framework accounts for the turbulent nature of atmospheric boundary layer flow by including artificial eddy viscosity in the network outputs, enhancing the model's ability to learn and accurately depict large-scale flow structures.
The reconstructed flow field and the effective wind speed are in good agreement with the actual data.
Furthermore, a transfer learning strategy is employed for the online deployment of pre-trained PINN, which requires less time than that of the actual physical flow.
This capability allows the framework to reconstruct wind flow fields in real time based on live data.
In the demo cases, the maximum error between the effective wind speed reconstructed online and the actual value at the wind turbine site is only 3.7\%.
The proposed data assimilation framework provides a universal tool for reconstructing spatiotemporal wind flow fields using various measurement data.
Additionally, it presents a viable approach for the online assimilation of real-time measurements.
To facilitate the utilization of wind energy, our framework's source code is openly accessible.

\end{abstract}

\keywords{Data assimilation \and Wind field reconstruction \and Physics-informed deep learning}

\section{Introduction}
Wind energy, recognized as a widely distributed renewable energy source, is the subject of extensive global research.
Wind turbines, specifically designed to harness this abundant resource, undergo the process of converting it into electrical energy.
This conversion plays a critical role in meeting the escalating requirements for electricity in both industrial production and daily life, and serves as a pivotal element in the pursuit of global carbon neutrality \cite{lienhard2023concurrent}.
The efficiency of energy capture from wind turbines is greatly influenced by factors such as design, site selection, and operational control, especially due to the complexity of turbulence within the atmospheric boundary layer \cite{doubrawa2019load,jin2023integration}.
To efficiently capture energy from the upstream inflow of wind turbines, wind speed measurement technologies are undergoing continuous evolution.
A prime example of this is light detection and ranging (LiDAR) \cite{harris2006lidar}, which has gained widespread adoption in recent years due to its capability to measure multi-point line-of-sight (LoS) wind speeds over a distance in a single direction \cite{gao2019investigation}.
Despite the advancements in technology, traditional single-point measurement methods remain prevalent, largely owing to their established reliability and lower cost \cite{jiang2024applicability}.
These methods include cup anemometers and wind vanes for velocity vector measurement, hot-wire anemometers or pitot tubes for velocity components assessment, as well as a variety of pressure sensors employed as auxiliary instruments \cite{mortensen1994wind}.
However, the existing measurement techniques can only partially characterize the wind flow field.
LiDAR, despite its ability to measure distant wind speeds, faces limitations in directly determining wind speed and direction, a challenge often termed the Cyclops’ dilemma \cite{dunne2011lidar, raach2014three}.
For single-point measurement methods, data measurement is restricted to a specific, individual location \cite{jiang2024applicability}.
Nonetheless, there is an urgent need for comprehensive wind flow field data for turbine design, site selection, and operational control.
Computational fluid dynamics (CFD) was employed to simulate the atmospheric boundary layer flow \cite{stoll2020large, tian2020numerical} and the aerodynamics of wind turbines \cite{bangga2021aerodynamic, rezaeiha2017effect}.
The complexity of wind fields, influenced by topography and vegetation, presents significant challenges for CFD simulations.
Consequently, the intricacies of flow in complex terrain are thoroughly examined in \cite{wenz2021impact, letzgus2021cfd, murali2017numerical}.
Furthermore, engineering models were utilized to enhance the predictive accuracy of high-fidelity CFD simulations for various design loads \cite{bangga2023utilizing}. 
Hand and Bolas \cite{hand2007blade} demonstrated that blade load mitigation control based on flow structures can significantly improve the effectiveness, reducing blade flap damage equivalent load by up to 30\%.

Benefiting from the rapid developments in data-driven science and machine learning, an increasing body of research focuses on estimating wind flow field states based on experimental measurements.
Despite the immense complexity of the atmospheric boundary layer flow, considerable efforts have been undertaken to capture the characteristics of the wind flow field through data-driven methodologies.
The unscented Kalman filter (UKF) was employed to determine the strength, location, and direction of advancing gusts from LiDAR measurements \cite{towers2016real}.
This examination encompassed a range of half-angles and look directions in LiDAR measurements, as well as diverse atmospheric stability conditions.
Such analysis facilitated a localized reconstruction of the flow field within the sectorial region spanned by the two laser beams.
A convolutional neural network (CNN) incorporating a wavelet transformer encoder was developed for probabilistic wind power forecasting \cite{wang2017deep}, effectively capturing uncertainties in wind power data.
A novel hybrid forecasting system, integrating mode decomposition with a recurrent neural network, was proposed for short-term wind speed prediction \cite{duan2021short}.
This hybrid model exhibited superior predictive performance compared to its counterparts based on long short-term memory networks (LSTM) and back-propagation neural networks.
Proper orthogonal decomposition was employed as an encoder in LSTM to predict key features of unsteady turbine wakes \cite{zhang2020novel}.
Indeed, these studies have successfully captured specific characteristics of the flow field with considerable accuracy.
Nevertheless, the reliance on simplified fluid dynamics models or the absence of physical models in these data-driven approaches limits their ability to extract further information from measurement data.
This limitation manifests in challenges related to accurately reconstructing flow fields over extensive areas and predicting their evolution.
Bauweraerts and Meyers \cite{bauweraerts2021reconstruction} employed a four-dimensional variational (4DVAR) data assimilation algorithm and large eddy simulation (LES) to reconstrut the wind flow field from LiDAR measurements. 
Mons et al. \cite{mons2021ensemble} adopted an ensemble-variational (EnVar) approach to infer corrections to LES subgrid models, resulting a data assimilation LES (DA-LES) method.
However, due to the nature of LES computations and their demands on processing power and time, the method of 4DVAR or EnVar integrated LES is limited to an offline mode.

Recently, physics-informed deep learning has gained significant attention in the field of fluid mechanics.
Originally developed for solving ordinary differential equations and partial differential equations (PDEs) \cite{lagaris1998artificial}, this method experienced a resurgence of interest with the advent of physics-informed neural networks (PINNs) \cite{raissi2019physics}.
The physical governing equations are integrated into the backpropagation neural network (BPNN).
Specifically, the residual of these equations is introduced as a normalization of the loss function of BPNN, with the gradient terms required for the computation of the residual being calculated through automatic differentiation \cite{baydin2018automatic}.
As the neural network is trained, the PINN gradually converges towards solving the physical equations by minimizing the loss function.
Raissi et al. \cite{raissi2019physics} integrated the Navier–Stokes equations into a deep neural network (DNN), using a two-dimensional (2D) incompressible laminar flow as a demonstration.
Here, the spatiotemporal flow fields were reconstructed from scattered data across the spatiotemporal domain.
Subsequent research \cite{raissi2020hidden} extended PINN to three-dimensional (3D) incompressible flow, inferring velocity and pressure fields directly from flow visualizations.
The PINN embedded with Euler equations, was also applied to learn compressible inviscid flows, capturing shock waves from a limited number of scattered points \cite{mao2020physics}.
Yan et al. \cite{yan2023exploring} demonstrated that PINN successfully learned hidden flow structures from sparse measurements.
The accuracy of these learned structures demonstrated high consistency with those obtained from high-fidelity flow field data, as analyzed through proper orthogonal decomposition.
Patel et al. \cite{patel2024turbulence} employed the Spalart-Allmaras (SA) turbulence model to PINNs for mean-flow reconstruction.
In the field of wind energy, pioneering efforts have been made in training PINN using Line-of-Sight (LoS) wind speed measurement data \cite{zhang2021spatiotemporal,zhang2021three}.
The trained PINN exhibited the capability to reconstruct detailed wind flow fields and, consequently, to assess effective wind speeds.
However, current applications of PINN in wind flow field analysis do not fully exploit the potential of various measurement data \cite{tian2024residual}.
Furthermore, these methods are limited to reconstructing the flow field for short periods in an offline mode and cannot utilize real-time measurement data \cite{zhang2021spatiotemporal, zhang2021three, cobelli2023physics}.

Therefore, a data assimilation framework that supports training PINN with mixed types of data is proposed in this paper.
The proposed framework supports the integration of various measurement data for training PINN, including LoS wind speed, velocity vector, velocity component, and pressure.
The trained PINN is capable of reconstructing the inflow flow field upstream of wind turbine sites.
Consequently, local characteristics such as effective speed can also be easily calculated based on this reconstruction.
Additionally, the introduction of transfer learning allows pre-trained PINN to adapt to real-time measurement data in an online mode.
The optimization objective of this data assimilation framework, specifically the loss function, consists of two components.
The first component of the loss function is the residual of the flow governing equations, which incorporates physical law constraints.
Considering the turbulent characteristics of atmospheric boundary layer flow, the parameterized N-S equations are employed as the governing flow equations.
The artificial viscosity is directly inferred by the network, eliminating the introduction of any turbulence models, which is elaborated in Section \ref{sec:2.1_PINN}.
The second component of the loss function depicts the discrepancy between the measurement data and the corresponding network output at identical locations, thereby enabling the assimilation of various types of measurement data.
This aspect is detailed in Section \ref{sec:2.2_Assimilation}.
The test case involves an atmospheric boundary layer flow simulated using SOWFA (Simulator fOr Wind Farm Applications) \cite{sowfa}.
The flow field within the horizontal plane upstream of the wind turbine site is chosen to be the test area of the proposed framework.
Detailed information on the SOWFA simulation setup and the hyperparameter settings for the data assimilation framework are provided in Section \ref{sec:3.1_Numerical setups}.
Section \ref{sec:3.2_SinVar} studies the impact of the locations of meteorological masts and the types of measurement data on the proposed data assimilation framework.
Section \ref{sec:3.3_Performance of data assimilation} investigates the impact of training with hybrid types of measurement data on accuracy.
The contribution levels of the two components of the loss function and their effects on accuracy are explored in Section \ref{sec:3.4_Influence of coef}.
Section \ref{sec:3.5_transfer_learning} assesses the feasibility of deploying this framework online through transfer learning.

This research addresses critical research gaps in the study of PINN and wind energy, particularly the previous limitations in utilizing mixed types of measurement data and weaknesses in online deployment.
To promote the more effective use of renewable energy, the proposed framework has been made open source. 
All source codes and test data are available at \url{https://github.com/Panda000001/WindAssimilation.git}

\section{Methodology}
\label{sec:2_Methodology}
\begin{figure*}[!htbp]
 \centering
 \includegraphics[width=\textwidth]{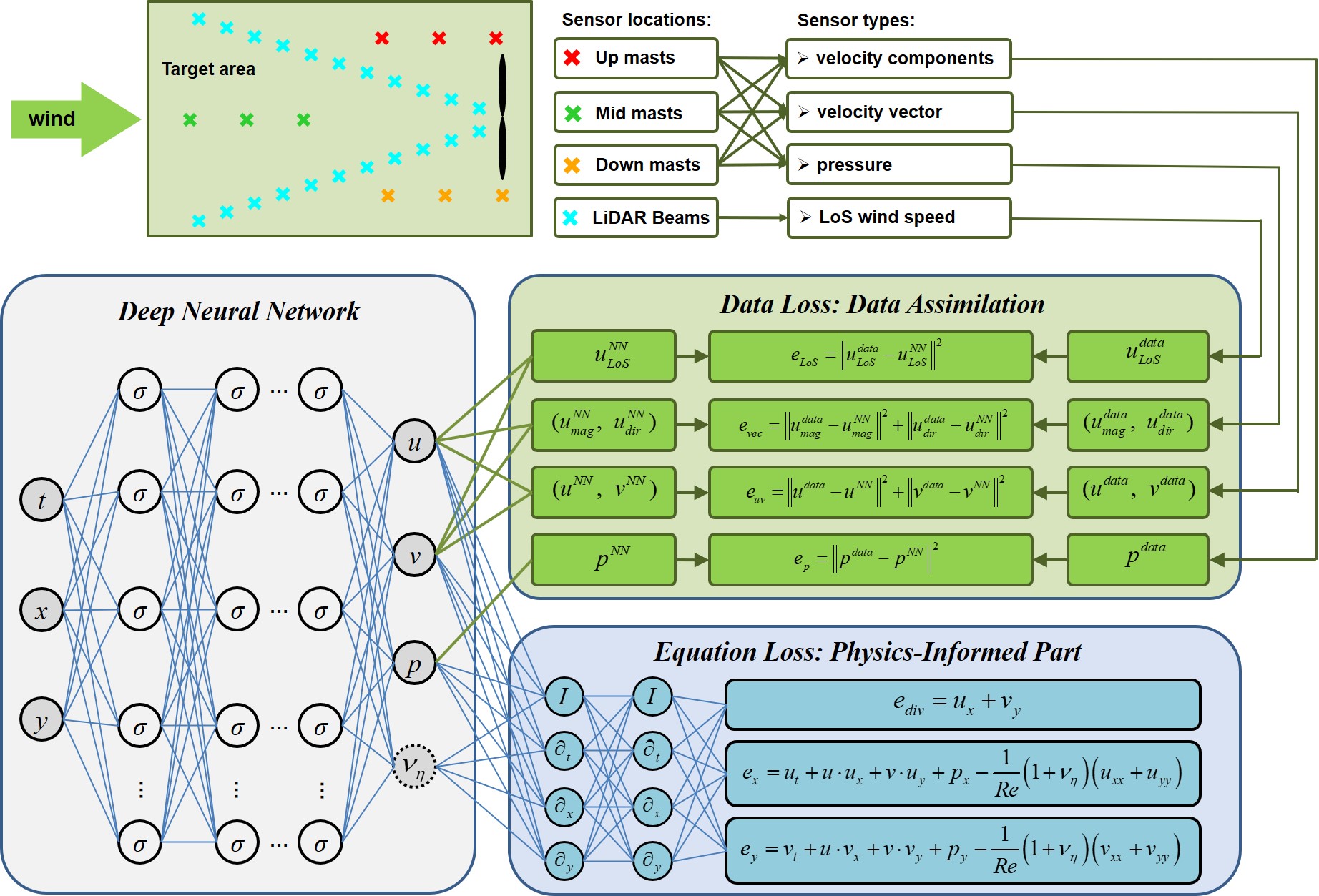}
 \caption{Schematic of the data assimilation framework for wind flow field. 
 The gray part shows the structure of the deep neural network, the blue part depicts the constraints of the flow governing equations, and the green part illustrates the data assimilation process.}
 \label{fig1:framework}
\end{figure*}

The data assimilation framework using PINN is illustrated in Figure \ref{fig1:framework}. 
The proposed framework is designed to reconstruct the spatiotemporal wind flow field from sparse data measured by various types of sensors.
At its core, the framework employs a deep neural network $\mathcal{N}$, which establishes the parametric mapping
\begin{equation}
{\mathcal{N}}: {\mathcal{X} \times \Theta}  \to {\mathcal{U}}
\label{eq:mapping}
\end{equation}
where $\mathcal{X}$ denotes the spatiotemporal coordinates $(t,x,y)$, $\Theta$ represents the set of all trainable parameters $\theta$ of $\mathcal{N}$, and $\mathcal{U}$ signifies the flow field variables $(u,v,p,\nu _{\eta})$, which represent the stream-wise velocity, the transverse velocity, the pressure and the artificial viscosity, respectively.
The loss function of $\mathcal{N}$ is composed of two components: 
\begin{equation}
{\mathcal{L}} = \lambda_{pde}\mathcal{L}_{pde} + \lambda_{data}\mathcal{L}_{data}
\label{eq:loss}
\end{equation}
where $\mathcal{L}_{pde}$, the equation loss, describes the extent to which the neural network's output conforms to the flow governing equations.
$\mathcal{L}_{data}$, the data loss, reflects the accuracy of the network's output in comparison to the actual measurement data.
The coefficients $\lambda_{pde}$ and $\lambda_{data}$ are used to balance these two components.
The specific compositions of $\mathcal{L}_{pde}$ and $\mathcal{L}_{data}$ are elaborated in Section \ref{sec:2.1_PINN} and Section \ref{sec:2.2_Assimilation}, respectively.
During the training process of the network, the objective of optimizing network parameters $\Theta$ is to minimize the loss function $\mathcal{L}$:
\begin{equation}
f(\Theta) = \arg \min \mathcal{L}\left( \Theta \right)
\label{eq:Minmize}
\end{equation}
The training set consists of measurement data from sparsely distributed locations within the spatial domain of the target area.
Details on the types of measurement data are provided in Section \ref{sec:2.2_Assimilation}.
Upon completion of training, the spatiotemporal evolution data across the entire target area can be reconstructed by the mapping defined in Equation (\ref{eq:mapping}). 

\subsection{\label{sec:2.1_PINN}Physics-informed neural network}
PINN was initially proposed in the literature \cite{raissi2019physics,raissi2020hidden} as a data-driven and mesh-free method to solve PDEs. 
As illustrated in Figure \ref{fig1:framework}, the basic structure of PINN is a DNN. 
The DNN receives temporal and spatial coordinates $(t,x,y)$ as inputs.
In its original form \cite{raissi2019physics}, PINN outputs the solution vector variables $(u,v,p)$ for the N-S equations.
However, it has been demonstrated that the accuracy of the original PINN decreases with increasing Reynolds number \cite{hanrahan2023studying,xu2022AMS}, while the wind fields typically involve high Reynolds number flows.
Therefore, in our framework, the PINN is constrained by the parameterized N-S equations rather than the original N-S equations, an approach validated in recent studies \cite{xu2021explore,xu2023spatiotemporal,zhang2021three}.
The parameterization of the viscosity coefficient in the N-S equations follows a methodology analogous to that of the Reynolds Averaged Navier-Stokes (RANS) equations.
The crux of the RANS approach lies in modeling the Reynolds stresses:
\begin{equation}
{\tau _{ij}} =  - \rho {\overline {u_i^{'}u_j^{'}}} 
\end{equation}
where ${\overline {(\cdot)}}$ denotes the Reynolds average.
Similar to Fick's law, $\rho {\overline {u_i^{'}u_j^{'}}}$ can be modeled as\cite{xu2021explore}:
\begin{equation}
 \rho \overline {u_i^{'}u_j^{'}}  = - \rho {\nu _t}(\frac{{\partial \overline {{u_i}} }}{{\partial {x_j}}} + \frac{{\partial \overline {{u_j}} }}{{\partial {x_i}}})
    \label{eq:eddy viscosity}
\end{equation}
where ${\nu _t}$ is the turbulence eddy viscosity, 
It is named as viscosity but different from the molecular viscosity in the original N-S equations.
A concept akin to this has been applied in LES, introducing sub-grid scale eddy viscosity ${\nu _{sgs}}$ to model the turbulent flows at subgrid scales.
While within the spectrum of hybrid RANS/LES approaches, ${\nu _t}$ in the control system embodies diverse significances across various regions of the computational domain.
As reviewed by Durbin \cite{DurbinAnnurev2018}, the concept of modeling turbulent transport through eddy viscosity forms the basis for much of the subsequent theoretical development.
This principle allows for the classical 2D N-S equations to be parameterized as follows:
\begin{align}
& {u_x} + {v_y} = 0\label{eq:para_continuity} \\
& {u_t} + u{u_x} + v{u_y} + {p_x} - \frac{1}{Re} (1+\nu _{\eta})({u_{xx}} + {u_{yy}}) = 0\label{eq:para_momentum_X} \\
& {v_t} + u{v_x} + v{v_y} + {p_y} - \frac{1}{Re} (1+\nu _{\eta})({v_{xx}} + {v_{yy}}) =0 \label{eq:para_momentum_y}
\end{align}
where $\nu _{\eta}$ represents an introduced artificial viscosity term for parameterization akin to the previously described method.
All variables are non-dimensional here. 
The reference length $D$ is the diameter of the wind turbine, the reference velocity ${U_\infty }$ is the average free stream velocity, and the reference pressure ${p_\infty }$ is $\rho {U_\infty }^2$.
The artificial eddy viscosity ${\nu _{\eta}}$ is non-dimensionalized by the molecular kinematic viscosity ${\nu} = {\mu}/{\rho}$, where ${\mu}$ is the dynamic viscosity and ${\rho}$ is the density. 
The Reynolds number $Re$ is defined as ${U_\infty}D/{\nu}$.

The undetermined parameter $\nu _{\eta}$, is directly inferred as an output variable of the DNN.
Thus, as shown in Figure \ref{fig1:framework}, the outputs of DNN are $(u,v,p, \nu _{\eta})$.
This method was originally proposed to explore missing flow dynamics \cite{xu2021explore}.
Consequently, the residual form of the non-dimensional 2D parameterized N-S equations are as follows,
\begin{align}
& {{e}_{div}} = {u_x} + {v_y} \label{eq:continuity} \\
& {{e}_{x}} = {u_t} + u{u_x} + v{u_y} + {p_x} - \frac{1}{Re} (1+\nu _{\eta})({u_{xx}} + {u_{yy}}) \label{eq:momentum_X} \\
& {{e}_{y}} = {v_t} + u{v_x} + v{v_y} + {p_y} - \frac{1}{Re} (1+\nu _{\eta})({v_{xx}} + {v_{yy}}) \label{eq:momentum_y}
\end{align}

The forward propagation in PINN is the same as that in a classical DNN, where $(u,v,p, \nu _{\eta})$ are computed from $(t,x,y)$.
As shown in Figure \ref{fig1:framework}, the input and output layers are interconnected by several hidden layers, each represented by a column.
The forward propagation process is expressed as follows:
\begin{align}
& {X^0} = {[t,x,y]^T} \label{eq:input_layer} \\
& {X^i} = \sigma \left( {{W^i}{X^{i - 1}} + {B^i}} \right),i \in [1,L - 1] \label{eq:hidden_layer} \\
& {X^L} = {{W^L}{X^{L - 1}} + {B^L}}={[u,v,p,{\nu _{\eta}}]^T} \label{eq:output_layer}
\end{align}
Here, $X^i$ denotes the values of the $i$-th layer, with $X^0$ and $X^L$ representing the input and output layers, respectively. 
The nonlinear activation function $\sigma$ of all hidden layers is the hyperbolic tangent $tanh$ in this study.
The trainable parameters ${W^i}$ and ${B^i}$ represent the weights and biases of the $i$-th layer, respectively. 
During the backward propagation phase in PINN, the derivatives of the outputs relative to the inputs are computed utilizing automatic differentiation \cite{baydin2018automatic}.
This allows for the calculation of the residuals of the governing equations (\ref{eq:continuity}), (\ref{eq:momentum_X}), and (\ref{eq:momentum_y}), leading to the determination of the equation loss $\mathcal{L}_{pde}$ in Equation (\ref{eq:loss}) as
\begin{equation}
\mathcal{L}_{pde} = {\left\|e_{div}\right\|}^2 + {\left\| e_x \right\|}^2 + {\left\| e_y \right\|}^2= \frac{1}{N_{eqns}}\sum\limits_{i=1}^{N_{eqns}} ({\left\| e_{div}(t^i,x^i,y^i) \right\|}^2 + {\left\| e_x(t^i,x^i,y^i) \right\|}^2 + {\left\| e_y(t^i,x^i,y^i) \right\|}^2)
 \label{eq:LossEqns}
\end{equation}
where ${N_{eqns}}$ is the number of equation points. 
The equation points can be arbitrarily selected across the spatial and temporal domain of the target area, making the PINN method a mesh-free approach.
Since the training process of the neural network parameters aims to minimize the loss function $\mathcal{L}$, as indicated in Equation (\ref{eq:Minmize}), incorporating the flow governing equations into $\mathcal{L}$ imposes a soft constraint of physical laws on the network training.

\subsection{\label{sec:2.2_Assimilation}Assimilating various kinds of measurements}
In the field of wind energy, experimental measurement is a crucial research method. 
However, existing measurement techniques typically capture only partial information about the flow field, such as velocity and pressure at certain discrete locations. 
As shown in the green part of Figure \ref{fig1:framework}, four types of data are studied in this paper.
It is noteworthy that the computational process of neural networks herein exclusively employs dimensionless quantities, with the reference values being identical to those delineated in Section \ref{sec:2.1_PINN}.
The first measurement technique is LiDAR \cite{harris2006lidar}, a method for measuring line-of-sight wind speeds in a specific direction, and it has emerged as a significant tool in recent years.
Due to the Cyclops' dilemma \cite{dunne2011lidar}, LiDAR can only measure a series of line-of-sight wind speeds along the laser beam, which is insufficient to fully characterize the entire flow field.
However, recent studies \cite{zhang2021spatiotemporal,zhang2021three} have demonstrated the potential of using PINN to reconstruct flow fields from LoS measurements. 
Therefore, LoS data is considered a vital source of experimental measurements for the proposed data assimilation framework.
To assimilate actual measured LoS wind speeds, the velocity components predicted by the DNN are first converted into predicted line-of-sight wind speeds, 
\begin{equation}
u_{LoS}^{NN}=u^{NN}cos\alpha-v^{NN}sin\alpha
 \label{eq:u_LoS_NN}
\end{equation}
where $\alpha$ is the angle of the LiDAR beam, $u^{NN}$ and $v^{NN}$ represent the velocity components along the $x$-axis and $y$-axis, respectively, as predicted by the DNN.
The $L_2$ norm of the difference between the actual measured LoS wind speeds $u_{LoS}^{data}$ and the DNN-predicted LoS wind speeds $u_{LoS}^{NN}$ is then incorporated as the first part of the data loss $\mathcal{L}_{data}$ in Equation (\ref{eq:loss}),
\begin{equation}
e_{LoS} = {\left\| {u_{LoS}^{data}-u_{LoS}^{NN}} \right\|}^2 = \frac{1}{N_{LoS}}\sum\limits_{i=1}^{N_{LoS}} {\left\| {u_{LoS}^{data}(t_i^{LoS},x_i^{LoS},y_i^{LoS})-u_{LoS}^{NN}(t_i^{LoS},x_i^{LoS},y_i^{LoS})} \right\|}^2
 \label{eq:LossLoS}
\end{equation}
where the $N_{LoS}$ is the number of LiDAR measurement points, and $(t_i^{LoS},x_i^{LoS},y_i^{LoS})$ are the temporal and spatial coordinates of these points.

The second category of measurement involves velocity vectors, typically obtained from instruments like cup anemometers, wind vanes, and helicoid propeller anemometers equipped with wind vanes. 
These devices measure either the velocity magnitude, the velocity direction, or both.
However, these sensors are usually contact-based, requiring installation within the wind field. 
Considering the economic costs, their large-scale deployment in practical engineering applications is challenging.
As a result, this category of measurement data is usually sparse, providing information from only a few locations.
To assimilate this type of actual measured data, the velocity components predicted by the DNN, $({u^{NN}},{v^{NN}})$, are converted into velocity vectors. The magnitude $u_{mag}^{NN}$ and direction $u_{dir}^{NN}$ of these vectors are calculated as follows:
\begin{equation}
(u_{mag}^{NN},u_{dir}^{NN}) = (\sqrt {{u^{N{N^2}}} + {v^{N{N^2}}}},arctan({v^{NN}}/{u^{NN}})
 \label{eq:u_vec_mag_NN}
\end{equation}
Subsequently, the $L_2$ norm of the discrepancy between the actual measured velocity vectors and those predicted by the DNN is incorporated as the second part of the data loss $\mathcal{L}_{data}$ in Equation (\ref{eq:loss}):
\begin{align} 
& e_{vec} = {e_{mag}} + {e_{dir}} \\
& e_{mag} = {\left\| {u_{mag}^{data} - u_{mag}^{NN}} \right\|^2} = \frac{1}{{{N_{vec}}}}\sum\limits_{i = 1}^{{N_{vec}}} {{{\left\| {u_{mag}^{data}(t_i^{vec},x_i^{vec},y_i^{vec}) - u_{mag}^{NN}(t_i^{vec},x_i^{vec},y_i^{vec})} \right\|}^2}} \\
& e_{dir} = {\left\| {u_{dir}^{data} - u_{dir}^{NN}} \right\|^2} = \frac{1}{{{N_{vec}}}}\sum\limits_{i = 1}^{{N_{vec}}} {{{\left\| {u_{dir}^{data}(t_i^{vec},x_i^{vec},y_i^{vec}) - u_{dir}^{NN}(t_i^{vec},x_i^{vec},y_i^{vec})} \right\|}^2}} 
 \label{eq:Loss_vec}
\end{align}
Here, $N_{vec}$ represents the number of velocity vector measurements, and $(t_i^{vec},x_i^{vec},y_i^{vec})$ are the spatiotemporal coordinates of the velocity vector sensors.
Additionally, $u_{mag}^{data}$ and $u_{dir}^{data}$ denote the actual measured magnitudes and directions of the wind speed, respectively.
Note that this is applicable when both magnitude and direction data are available. 
If only magnitude or direction data is present, then only $e_{mag}$ or $e_{dir}$ should be calculated, respectively.

The third type of measurement data involves the velocity components, either aligned with or perpendicular to the inflow direction.
Instruments such as hot-wire anemometers, tube anemometers, or ultrasonic anemometers typically gather this data.
Since velocity components are directly predicted by the DNN,  the $L_2$ norm of the difference between the actual measured velocity components and the prediction of the DNN is directly included as the third part of the data loss $\mathcal{L}_{data}$ in Equation (\ref{eq:loss}),
\begin{align} 
& e_{comp} = {e_{u}} + {e_{v}} \\
& e_{u} = {\left\| {u^{data} - u^{NN}} \right\|^2} = \frac{1}{{{N_{comp}}}}\sum\limits_{i = 1}^{{N_{comp}}} {{{\left\| {u^{data}(t_i^{comp},x_i^{comp},y_i^{comp}) - u^{NN}(t_i^{comp},x_i^{comp},y_i^{comp})} \right\|}^2}} \\
& e_{v} = {\left\| {v^{data} - v^{NN}} \right\|^2} = \frac{1}{{{N_{comp}}}}\sum\limits_{i = 1}^{{N_{comp}}} {{{\left\| {v^{data}(t_i^{comp},x_i^{comp},y_i^{comp}) - v^{NN}(t_i^{comp},x_i^{comp},y_i^{comp})} \right\|}^2}} 
 \label{eq:Loss_uv}
\end{align}
where $N_{comp}$ denotes the number of velocity component data points, $(t_i^{comp},x_i^{comp},y_i^{comp})$ are the spatiotemporal coordinates of the velocity component sensors, $u^{data}$ and $v^{data}$ represent the actual measured velocity components along the $x$-axis and $y$-axis, respectively.
Note that this is applicable when both streamwise and transverse measurement data are available. 
If only one component, either streamwise or transverse data, is present, then only  $e_u$ or $e_v$ should be calculated, respectively.

The fourth type of measurement data considered in this study is pressure $p$.
Pressure measurements are commonly performed using a variety of pressure sensors, including piezoresistive, capacitive, and piezoelectric types.
These measurements are often used as a supplementary means in wind field measurements, conducted at a limited number of locations. 
The data from pressure sensors are also incorporated into the proposed data assimilation framework. 
In our framework, the $L_2$ norm of the discrepancy between the actual measured pressure and the pressure predicted by the DNN constitutes the fourth part of the data loss  $\mathcal{L}_{data}$ in Equation (\ref{eq:loss}),
\begin{equation}
e_p = {\left\| {p^{data}-p^{NN}} \right\|}^2 = \frac{1}{N_{p}}\sum\limits_{i=1}^{N_{p}} {\left\| {p^{data}(t_i^p,x_i^p,y_i^p)-p^{NN}(t_i^p,x_i^p,y_i^p)} \right\|}^2
 \label{eq:LossP}
\end{equation}
where the $N_p$ represents the number of pressure sensors, $p^{data}(t_i^p,x_i^p,y_i^p)$ is the actual pressure data measured by the pressure sensor at location $(x_i^p,y_i^p)$ and time $t_i^p$, and $p_i^{NN}(t_i^p,x_i^p,y_i^p)$ is the pressure data predicted by the DNN at the same location and time.

The methods described above represent common measurement techniques in wind engineering.
When all these types of measurement data are available, the data loss $\mathcal{L}_{data}$ in Equation (\ref{eq:loss}) is formulated as,
\begin{equation}
\mathcal{L}_{data} = e_p + e_{LoS} + e_{vec} + e_{comp}
\label{eq:loss_data}
\end{equation}
In this equation, each term of $e_p$, $e_{LoS}$, $e_{vec}$, and $e_{comp}$ corresponds to the data loss component associated with each type of measurement.
When only a subset of the measurement data is available, only the relevant components are calculated.
This approach allows for flexibility in the data assimilation process, accommodating the availability of different types of measurement data.
Furthermore, this framework allows for the integration of boundary conditions and initial conditions if available, by discretizing them into data points and incorporating them as soft constraints within the loss function.
In addition, to evaluate the accuracy of the reconstructed flow field, all four types of sensor measurement data utilized for training in this work are derived from the numerical results generated by SOWFA.
Consequently, all measurement data mentioned in this paper are indicative of data extracted from numerical simulations using virtual sensors.

\section{Results and discussion}
\label{sec:3_Results and discussion}
This section evaluates the framework for assimilating wind field data as previously outlined.
The LES wind farm simulator, SOWFA \cite{sowfa}, is employed to generate wind field data, which serves as a ground truth for validating the accuracy of the reconstructed flow fields.
Measurement points are strategically placed at various locations upstream of the wind turbine installation site to gather pressure and velocity data.
The four types of measurement data discussed in Section \ref{sec:2.2_Assimilation} are assimilated either partially or entirely by the PINN. 
The trained PINN is subsequently employed to reconstruct the complete velocity fields of the target area.
Section \ref{sec:3.1_Numerical setups} provides the specific parameters used for wind field simulation and network training. 
The impact of the locations of meteorological masts and the types of measurement data is studied in Section \ref{sec:3.2_SinVar}.
The accuracy of assimilating hybrid types of measurement data is assessed in Section \ref{sec:3.3_Performance of data assimilation}, and the influence of the weights in the loss function is analyzed in Section \ref{sec:3.4_Influence of coef}.
Section \ref{sec:3.5_transfer_learning} explores the application of transfer learning for the online deployment of the pre-trained framework to assimilate real-time measurements.

\subsection{\label{sec:3.1_Numerical setups}Numerical setups}
\begin{figure*}[!htbp]
\centering
  \subfloat[Simulation domain]{%
    \includegraphics[width=0.4\textwidth]{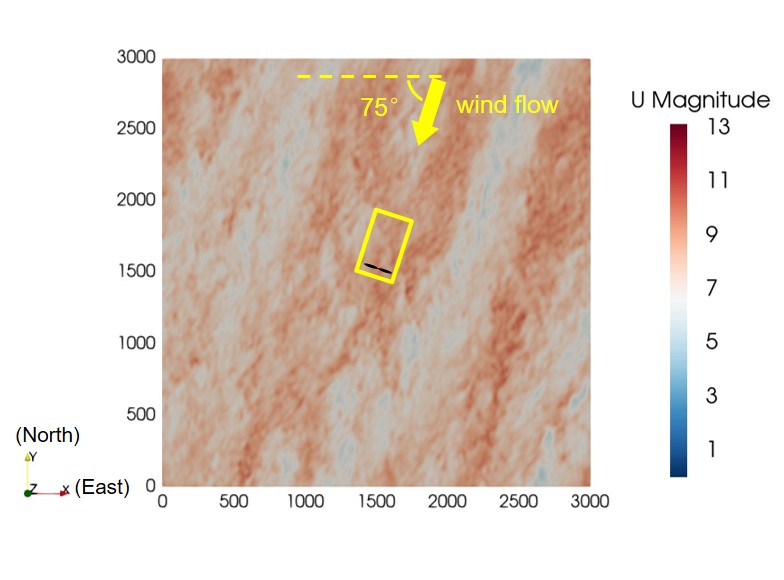}
    \label{fig2a:CFD_domain}
  }
  \hfill
  \subfloat[Target area and measurements (in sub-coordinate system)]{%
    \raisebox{0.02\textwidth}{\includegraphics[width=0.55\textwidth]{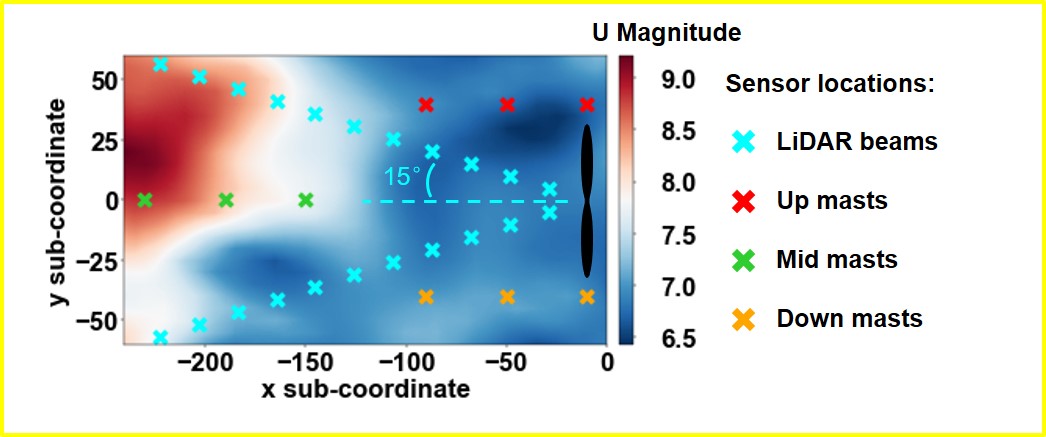}}
    \label{fig2b:PINN_domain}
  }
\caption{Velocity magnitude ($m/s$) at the turbine hub height $H=90m$ and the arrangement of sensors in target area.}
\label{fig2:CFD_Result}
\end{figure*}

SOWFA, a LES wind farm simulator developed by the National Renewable Energy Laboratory (NREL) and based on OpenFOAM, is utilized in this study.
It is specifically designed for simulating the turbulent atmospheric boundary layer.
For our research, SOWFA generates a wind field over a 3 $km$ $\times $ 3 $km$ $\times $ 1 $km$ cuboid area with flat terrain.
The grid generation and computational settings follow the guidelines recommended in the literature \cite{churchfield2012numerical, zhang2021spatiotemporal, zhang2021three}.
A uniform grid of 12 $m$ $\times $ 12 $m$ $\times $ 12 $m$ is used, resulting in a total of $5.25\times10^6$ cells. 
The average freestream wind speed is set at 8 $m/s$ and the atmospheric stability is neutral.
The simulation begins with a time step of 0.5 $s$ for 20,000 $s$ to establish a quasi-equilibrium flow field.
Subsequently, the simulation continues for 500 $s$ with a time step of 0.02 $s$, where the flow field from the last 100 $s$ is used as the ground truth. 
While the blockage effect of the wind turbine rotor can disturb the flow field near the turbine \cite{medici2011upstream}, this effect is negligible in the free-stream flow of the upstream target area \cite{towers2016real}.
Therefore, the aerodynamic aspects of the wind turbine are not considered in this study.
The center of the computational domain is designated as the turbine installation site, with a rotor diameter $D$ = 60 $m$. 
In this configuration, the Reynolds number is $4.8 \times 10^{7}$ based on the reference diameter $D$ = 60 $m$.
Figure \ref{fig2a:CFD_domain} illustrates the area used for the data assimilation study.
The target area for this study is a rectangular region measuring $4D \times 2D$ within the horizontal plane at a hub height $H_{hub}$ of 90 $m$ upstream of the wind turbine.
For clarity, the target area is represented in a relative coordinate system as depicted in Figure \ref{fig2b:PINN_domain}, with the rotor axis direction as the $x$-axis, the inflow direction as the $+x$ direction, and the rotor center in the relative coordinate system located at (-10 $m$, 0 $m$).

Data for training the PINN are collected using virtual sensors arranged in the target rectangular area during the final 100 $s$ of the simulation.
The data collection process involves projecting the simulated data onto the positions of these virtual sensors.
As depicted in Figure \ref{fig2b:PINN_domain}, the setup includes two laser beams with a 30$^\circ$ included angle for measuring LoS data.
The two beams are symmetrically positioned along the rotor axis, extend a range of 220 $m$, and feature a resolution of 20 $m$.
Consequently, this configuration results in 11 measurement points on each beam.
Following the pioneering research by Zhang and Zhao \cite{zhang2021spatiotemporal}, the LoS wind speed measurements are relatively convenient to obtain.
Therefore, the PINN model trained exclusively with LoS wind speed data is established as the baseline case.
Additional three types of data, velocity vector, velocity components and pressure are subsequently incorporated into the training process of the PINN model, alongside the LoS wind speed data, in order to evaluate the performance of the proposed data assimilation framework.
subsequently, additional types of data are integrated with LoS wind speed data to train the PINN, aiming to assess the performance of the proposed data assimilation framework. 

Three common types of measurement data in wind farms, including velocity vector, velocity components, and pressure, are individually or collectively incorporated into the training dataset alongside the LoS wind speed data to train the PINN model.
Following the convention of meteorological masts commonly employed in wind farms, additional data is measured at the positions of three sets of virtual meteorological masts, denoted as "Up", "Mid", and "Down", as illustrated in Figure \ref{fig2b:PINN_domain}.
Theoretically, meteorological masts can be placed at any location within the target area. 
Considering practical operability in real-world applications, we selected the three sets of positions, each featuring three equidistant measurement points.
For the set of "Up" masts, three sensors are arranged at $y$ = 40 $m$, with their $x$ coordinates uniformly spanning from -90 $m$ to -10 $m$.
Similarly, for the "Mid" set, the three sensors are positioned along the rotor axis at $y$ = 0 $m$, with their $x$ coordinates uniformly ranging from -230 $m$ to -150 $m$.
Three sensors in "Down" set are placed at $y$ = -40 $m$, with their $x$ coordinates evenly distributed within the range of -90 $m$ to -10 $m$.
In this paper, all of four measurement methods have a sampling frequency of 1 $Hz$, equating to one data point per second.
This uniform sampling rate simplifies the data collection process. 
In practical scenarios, both sensor positions and sampling frequencies can be varied as needed.
Within this sampling configuration, the number of LoS wind speed data points utilized for training the PINN amounts to $N_{LoS}$=2200, in addition to additional $N_p$, $N_{comp}$, or$N_{vec}$ =300 data points.

\begin{figure*}[!htbp]
 \centering
 \includegraphics[width=0.6\textwidth]{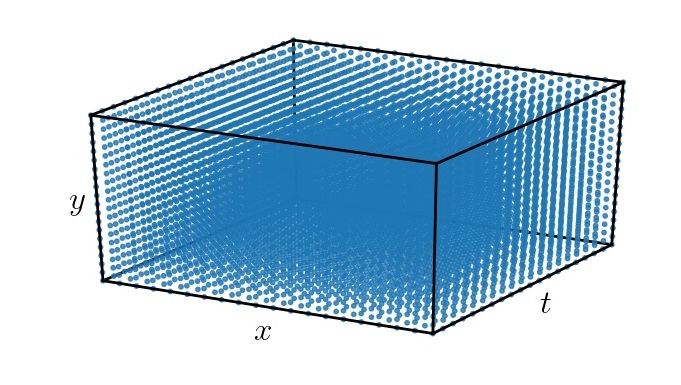}
 \caption{A schematic of uniformly generated equation points. During the training of the neural network, these equation points offer solely spatiotemporal coordinate values without corresponding values of the flow field.}
 \label{figB:App_Equation_Points}
\end{figure*}

Equation points consist of spatiotemporal coordinates uniformly generated within the rectangular target area depicted in Figure \ref{fig2b:PINN_domain} and over a duration of 100 seconds.
Sampling is conducted on the $x$, $y$, and $t$ axes with 81, 41, and 100 points respectively, culminating in a total of 332,100 equation points.
The uniformly generated equation points are depicted in Figure \ref{figB:App_Equation_Points}, they are a set of spatiotemporal coordinates without flow field data.
In the training process of the neural network, these equation points offer spatiotemporal coordinate values to calculate the residuals of flow governing equations at these discrete positions.

The proposed framework, based on a neural network, involves several hyperparameters that determine its structure and training configuration.
To focus exclusively on the impact of different data types on data assimilation, these hyperparameters are uniformly set across all eight cases.
Table \ref{tab:hyper} lists the configuration of the hyperparameters utilized in this study, along with their respective meanings.
The equation points in each batch during training
$N_{eqns}$ is the number of equation points in each batch during training.
In particular, $\lambda_{pde}$ and $\lambda_{data}$ are distinctive parameters of the PINN, signifying the weight coefficients for equation loss $\mathcal{L}_{pde}$ and data loss $\mathcal{L}_{data}$, respectively.
In Section \ref{sec:3.2_SinVar} and Section \ref{sec:3.3_Performance of data assimilation}, adhering to the settings of Zhang and Zhao \cite{zhang2021spatiotemporal}, $\lambda_{pde}$ and $\lambda_{data}$ are both set to a value of 1.
In Section \ref{sec:3.4_Influence of coef}, the influence of the weight hyperparameters $\lambda_{pde}$ and $\lambda_{data}$ will be further explored.
To eliminate the impact of randomness inherent in the neural network training process, each case is independently trained five times.
All training sessions are executed on the NVIDIA Tesla V100 GPU. 
The average duration of each training step is approximately 0.1 seconds, culminating in a total training time of around 2.8 hours for each case.
Once training is completed, reconstructing the complete flow field for a single time instance takes merely 0.002 seconds.

\begin{table}[!htbp]
    \centering
    \caption{The hyperparameters of PINN structure and training setup.}
    \renewcommand{\arraystretch}{1.5}
    \begin{tabular}{ccc}
    \toprule
     Hyperparameter & Meaning & Value \\ \midrule
     $L$ & Neural network layers & 11 \\
     $N_{h}$  & Neurons per hidden layers & 128\\
     $\sigma$  & Activation functions that introduce non-linearity & $tanh$\\
     $N_{eqns}$ & Number of randomly selected equation points in one training step & 1000\\
     $epochs$ & Total number of training steps & $10^6$\\
     $lr$ & Learning rate, i.e., the step size in gradient descent & $10^{-4}$ \\
     $optimizer$ & Algorithm for minimizing the loss function & $Adam$\\ 
     $\lambda_{pde}$ & Weight of equation loss & 1\\ 
     $\lambda_{data}$ & Weight of data loss & 1\\ \bottomrule
    \end{tabular}
    \label{tab:hyper}
\end{table}

Upon completion of the training, the trained model is employed to reconstruct the unsteady flow field within the specified target area illustrated in Figure \ref{fig2b:PINN_domain}, thereby achieving the objective of assimilating various types of measured data.
To quantitatively evaluate the accuracy of the framework in assimilating the flow field, the root-mean-squared error ($RMSE$) is used as a metric, following the approach in Zhang and Zhao \cite{zhang2021spatiotemporal}. 
The $RMSE$ is defined as follows:
\begin{align} 
& RMS{E_{mag}} = \frac{1}{{{N_t}}}\sum\limits_{t = 1}^{{N_t}} {\sqrt {\frac{1}{{{N_s}}}\sum\limits_{i = 1}^{{N_s}} {\left[ {u_{mag}^{true}({x_i},{y_i},t) - u_{mag}^{NN}({x_i},{y_i},t)} \right]^2} } } \\
& RMS{E_{dir}} = \frac{1}{{{N_t}}}\sum\limits_{t = 1}^{{N_t}} {\sqrt {\frac{1}{{{N_s}}}\sum\limits_{i = 1}^{{N_s}} {\left[ {u_{dir}^{true}({x_i},{y_i},t) - u_{dir}^{NN}({x_i},{y_i},t)} \right]^2} } }
 \label{eq:MRMSE}
\end{align}
where $N_t$ represents the total number of time steps, which is 100 in this work, and $N_s$ denotes the total number of spatial points.
The variables utilized for calculating the $RMSE$, as well as those employed in subsequent visualizations, are dimensional variables that have been restored to their original units. 
Dimensionless quantities are exclusively used during the neural network training process.
For this study, a $N_x \times N_y = 49 \times 25$ uniform grid of the target area illustrated in Figure \ref{fig2b:PINN_domain} and 100 time steps within 100 $s$ are used for assessment, resulting in $N_s = N_x \times N_y = 1225$ and $N_t=100$. 
The superscript $NN$ and $true$ on $u$ indicate the network-reconstructed values and ground truth values simulated by SOWFA, respectively, while the subscript $mag$ and $dir$ represent magnitude and direction.
This metric is applied to evaluate the discrepancies in both velocity magnitude and velocity direction compared to the ground truth. 
Due to the inherent randomness in neural network training, each case in this paper is independently trained five times.
In this paper, the data used for training the neural network are all dimensionless, with dimensions only restored during the visualization of flow fields and the evaluation of $RMSE$.

\subsection{\label{sec:3.2_SinVar}The influence of positions and types of additional data}
\begin{figure*}[!htbp]
 \centering
 \includegraphics[width=\textwidth]{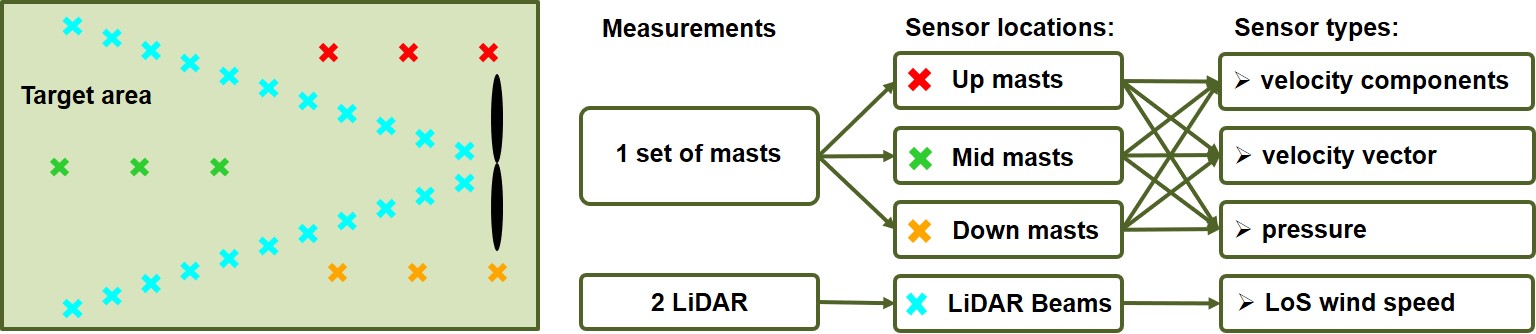}
 \caption{A schematic illustration of data measurement configurations utilized for investigating the effects of measurement location and data type.
 Basic LoS wind speed measurements, along with measurements from a single data type on a set of meteorological masts, are collected.}
 \label{figB:App_IFrange_Sensor}
\end{figure*}

In this section, we explore the impact of the locations of meteorological masts and the types of additional measurement data on the proposed data assimilation framework. 
A single type of additional data measured at each set of virtual meteorological mast locations is employed in conjunction with LoS wind speed data for training, compared to a baseline case which solely employs LoS wind speed data for training. 
Figure \ref{figB:App_IFrange_Sensor} illustrates the locations of three sets of virtual meteorological mast, along with the types of measured data at these locations.
Table \ref{tab:SinVar} presents the types of training data and their measurement locations used in three sets of numerical experiments (Up, Mid, Down) aimed at investigating the impact of location, and three sets of numerical experiments (p, comp, vec) designed to study the impact of data type. 
The baseline case uses only LoS wind speed data, whereas each of the other cases incorporates an additional single type of measurement data from a set of virtual meteorological masts.
To mitigate the inherent randomness in neural network training, each case is independently trained five times.

\begin{table}[!htbp]
    \centering
    \caption{The numerical experiment configurations for the study of the influence of positions and types of additional data.}
    \renewcommand{\arraystretch}{1.5}
    \begin{tabular}{ccccc}
    \toprule
     Group & Baseline data & Additional data type & Additional data position & Additional data points \\ \midrule
     Baseline & LoS & None & None & $N_p$=0,$N_{comp}$=0,$N_{vec}$=0 \\
     \hline
     \multirow{3}{*}{Up} & LoS & $u_{vec}$ & 3 Up masts & $N_p$=0,$N_{comp}$=0,$N_{vec}$=300 \\
                         & LoS & $u_{comp}$ & 3 Up masts & $N_p$=0,$N_{comp}$=300,$N_{vec}$=0 \\
                         & LoS & $p$ & 3 Up masts & $N_p$=300,$N_{comp}$=0,$N_{vec}$=0 \\
     \hline
     \multirow{3}{*}{Mid} & LoS & $u_{vec}$ & 3 Mid masts & $N_p$=0,$N_{comp}$=0,$N_{vec}$=300 \\
                          & LoS & $u_{comp}$ & 3 Mid masts & $N_p$=0,$N_{comp}$=300,$N_{vec}$=0 \\
                          & LoS & $p$ & 3 Mid masts & $N_p$=300,$N_{comp}$=0,$N_{vec}$=0 \\
     \hline
     \multirow{3}{*}{Down} & LoS & $u_{vec}$ & 3 Down masts & $N_p$=0,$N_{comp}$=0,$N_{vec}$=300 \\
                           & LoS & $u_{comp}$ & 3 Down masts & $N_p$=0,$N_{comp}$=300,$N_{vec}$=0 \\
                           & LoS & $p$ & 3 Down masts & $N_p$=300,$N_{comp}$=0,$N_{vec}$=0 \\
     \hline
     \multirow{3}{*}{p} & LoS & $p$ & 3 Up masts & $N_p$=300,$N_{comp}$=0,$N_{vec}$=0 \\
                        & LoS & $p$ & 3 Mid masts & $N_p$=300,$N_{comp}$=0,$N_{vec}$=0 \\
                        & LoS & $p$ & 3 Down masts& $N_p$=300,$N_{comp}$=0,$N_{vec}$=0 \\
     \hline
     \multirow{3}{*}{comp} & LoS & $u_{comp}$ & 3 Up masts & $N_p$=0,$N_{comp}$=300,$N_{vec}$=0 \\
                           & LoS & $u_{comp}$ & 3 Mid masts & $N_p$=0,$N_{comp}$=300,$N_{vec}$=0 \\
                           & LoS & $u_{comp}$ & 3 Down masts & $N_p$=0,$N_{comp}$=300,$N_{vec}$=0 \\
     \hline
     \multirow{3}{*}{vec} & LoS & $u_{vec}$ & 3 Up masts & $N_p$=0,$N_{comp}$=0,$N_{vec}$=300 \\
                          & LoS & $u_{vec}$ & 3 Mid masts & $N_p$=0,$N_{comp}$=0,$N_{vec}$=300 \\
                          & LoS & $u_{vec}$ & 3 Down masts & $N_p$=0,$N_{comp}$=0,$N_{vec}$=300 \\ \bottomrule
    \end{tabular}
    \label{tab:SinVar}
\end{table}

\begin{figure*}[!htbp]
\centering
  \subfloat[$RMSE$ of Wind speed magnitude]{
    \includegraphics[width=0.48\textwidth]{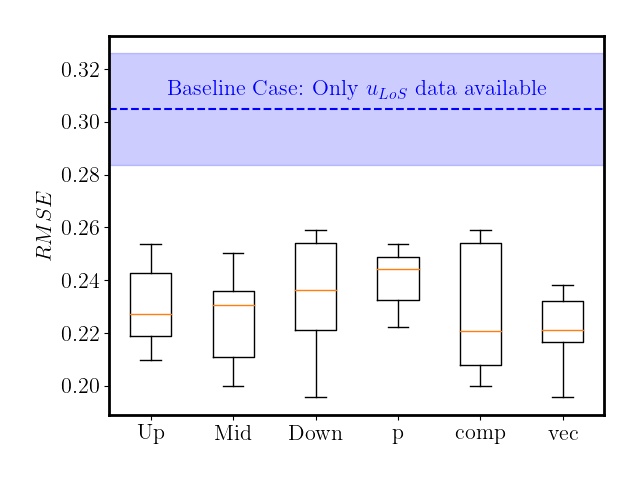} \label{figBa:IFrange_MRSE_mag} }
  \subfloat[$RMSE$ of Wind speed direction]{
    \includegraphics[width=0.48\textwidth]{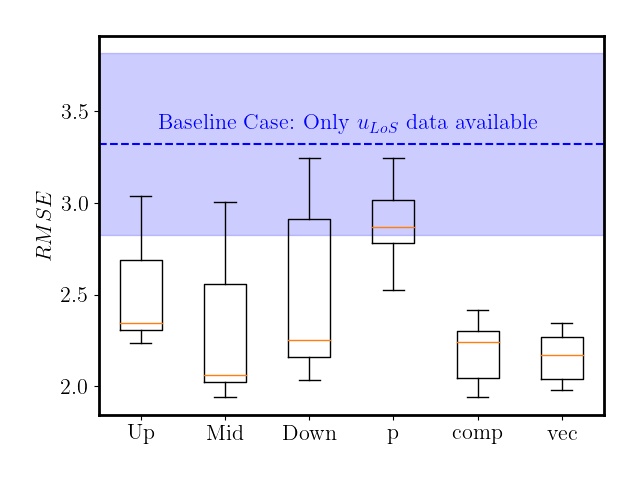} \label{figBb:IFrange_MRSE_dir} }
\caption{The influence of positions and types of additional data to $RMSE$. The $RMSE$ values represent the relative root mean square errors between the reconstructed wind flow fields generated by the trained PINN and the ground truth computed by SOWFA. The impact range of three sets of meteorological mast positions (each set comprising three measurement locations) and three types of data is illustrated. Box plots labeled "Up/Mid/Down" depict the results obtained from placing each type of data at the corresponding positions of the three sets of meteorological masts, each tested in five independent experiments, resulting in a total of 15 numerical experiments in one group. Similarly, box plots labeled "p/comp/vec" represent the results obtained from employing each type of measurement method at the three sets of meteorological mast positions, again tested in five independent experiments, resulting in a total of 15 numerical experiments in one group. The baseline case utilizes only LoS data, and five independent experiments are conducted. The mean and standard deviation are represented by the blue dashed lines and shaded areas, respectively.}
\label{figB:IFrange_MRSE}
\end{figure*}

Figure \ref{figB:IFrange_MRSE} employs box plots to display the range of impacts that measurement location and data type have on the $RMSE$ of reconstructed flow fields.
It presents the effects of three meteorological mast locations and three data types.
The box plots labeled Up/Mid/Down show the measurements of the three data types at each location, based on fifteen numerical experiments from five independent trials per case. 
Similarly, the box plots for p/comp/vec display each measurement type at the three set of meteorological mast locations, including a total of fifteen results from five independent experiments per case.
In the box plots, the whiskers represent the minimum and maximum values within 1.5 times the interquartile range (IQR) from the first and third quartiles, with the box edges indicating the first (Q1) and third (Q3) quartiles, and the orange line within the box showing the median.
The results indicate that the type of data has a more pronounced impact than the location.
Among the three data types, the vec group exhibits the lowest median $RMSE$ for both wind speed magnitude and direction, while the p group shows the highest $RMSE$.
In addition, the comp group's $RMSE$ being slightly higher than that of the vec group but still close.
From the perspective of location, the impact is more complex, with the $RMSE$ range for wind speed magnitude being similar across the Up/Mid/Down groups, but the Mid group showing a better trend for both wind speed magnitude and direction $RMSE$.

In conclusion, under the current settings of positions, placing sensors in the upstream Mid group is more beneficial for data assimilation.
The contribution of velocity vectors is close to and slightly higher than that of velocity components, both of which are more significant than the contribution of pressure data.
However, these are trend-based conclusions derived from the numerical experiments and may not strictly apply to individual cases.
Nevertheless, it is also important to highlight that all cases involving the assimilation of extra data types demonstrated a noticeable improvement compared to the baseline case, which relied exclusively on LoS wind speed data.
This observation unequivocally supports the effectiveness of the proposed data assimilation framework.

\subsection{\label{sec:3.3_Performance of data assimilation} Performance of assimilating hybrid types of data}
\begin{figure*}[!htbp]
 \centering
 \includegraphics[width=0.85\textwidth]{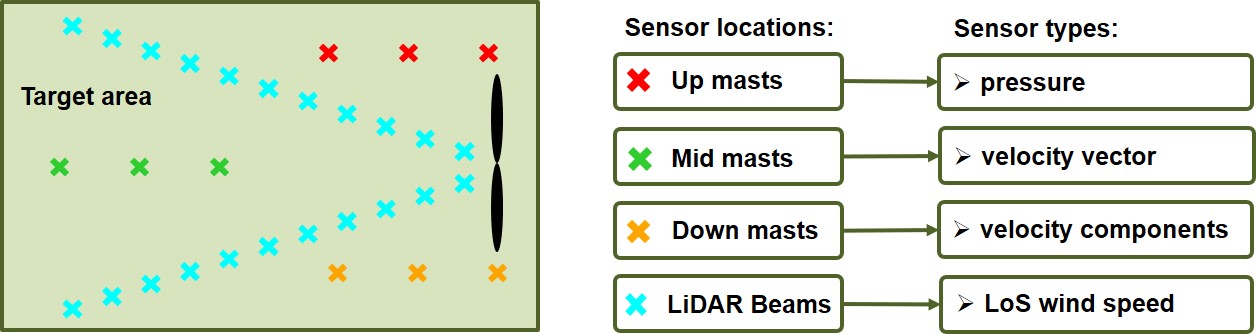}
 \caption{A schematic of data measurement configurations used for studying the assimilation of hybrid types of data.
 Basic LoS wind speed measurements, along with measurements from one to three sets of meteorological masts, are collected.
 The Up masts measures pressure, the Mid masts measures velocity vectors, and the Down masts measures velocity components.}
 \label{fig:App_hybrid_Sensor}
\end{figure*}

To evaluate the performance in assimilating hybrid types of data, four measurement data types are combined into eight distinct combinations for training the PINN.
The specific data types and the number of data points used in these eight cases are detailed in Table \ref{tab:case}.
The sensor locations and corresponding types are illustrated in Figure \ref{fig:App_hybrid_Sensor}.
As the LoS wind speed measurements are relatively convenient to obtain, Case 1, which involves training the PINN exclusively with LoS wind speed data, is established as the baseline case.
Cases 2 to 4 each add three measurement points of a different type to Case 1: three velocity vector points, three velocity component points, and three pressure points, respectively.
Cases 5 to 7 incorporate two other types of measurement data in each case, while Case 8 employs all four types of data.

\begin{table}[!htbp]
    \centering
    \caption{The numerical experiment configurations for the study of assimilating hybrid types of data.}
    \renewcommand{\arraystretch}{1.5}
    \begin{tabular}{lcc}
    \toprule
     Case  & Additional data type & Additional data points \\ \midrule
     Case 1: LoS & None & $N_p$=0,$N_{comp}$=0,$N_{vec}$=0 \\
     Case 2: LoS+vec & $u_{vec}$ & $N_p$=0,$N_{comp}$=0,$N_{vec}$=300\\
     Case 3: LoS+comp & $u_{comp}$ & $N_p$=0,$N_{comp}$=300,$N_{vec}$=0\\
     Case 4: LoS+p & $p$ & $N_p$=300,$N_{comp}$=0,$N_{vec}$=0\\
     Case 5: LoS+vec+comp & $u_{vec}$, $u_{comp}$ & $N_p$=0,$N_{comp}$=300,$N_{vec}$=300\\
     Case 6: LoS+vec+p & $u_{vec}$, $p$ & $N_p$=300,$N_{comp}$=0,$N_{vec}$=300\\
     Case 7: LoS+comp+p & $u_{comp}$, $p$ & $N_p$=300,$N_{comp}$=300,$N_{vec}$=0\\
     Case 8: LoS+vec+comp+p & $u_{vec}$, $u_{comp}$, $p$ & $N_p$=300,$N_{comp}$=300,$N_{vec}$=300\\ \bottomrule
    \end{tabular}
    \label{tab:case}
\end{table}

\begin{figure*}[!htbp]
\centering
  \subfloat[Wind speed magnitude]{
    \includegraphics[width=0.4\textwidth]{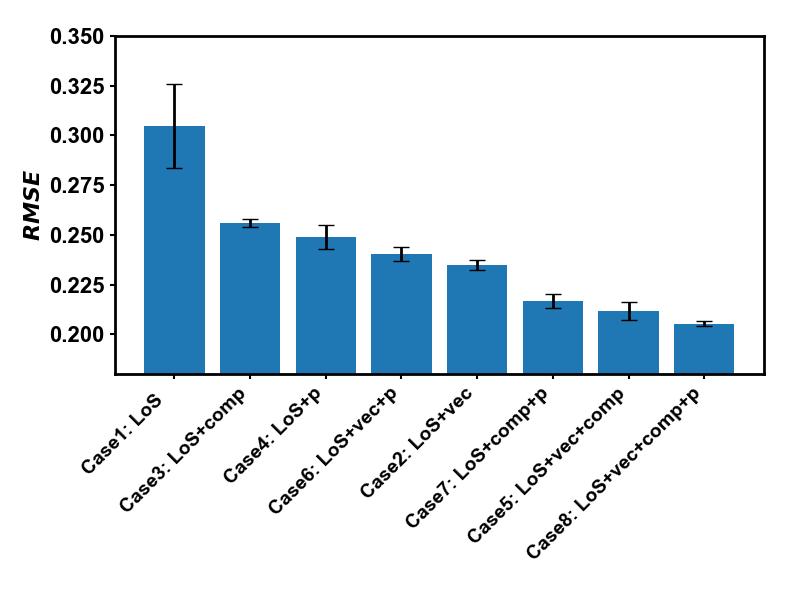} \label{fig3a:Assi_MRSE_mag} }
  \subfloat[Wind speed direction]{
    \includegraphics[width=0.4\textwidth]{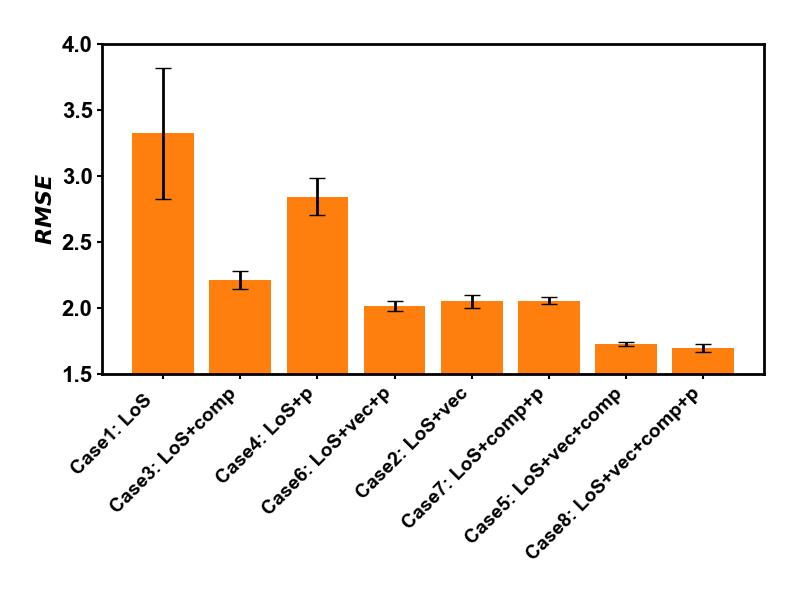} \label{fig3b:Assi_MRSE_dir} }
\caption{The average $RMSE$ between the reconstructed wind flow fields by PINN and the ground truth within 100$s$. The $RMSE$ value is the average result of five independent trainings, with the error bars representing the standard deviation of the five trainings.}
\label{fig3:Assi_MRSE}
\end{figure*}

In Case 1, the PINN is trained exclusively with LoS wind speed data, following the same settings used in Zhang and Zhao \cite{zhang2021spatiotemporal}.
The results of Case 1 establish a baseline for comparison in this study.
Building on Case 1, the other three types of measurement data are either individually or collectively incorporated into the training set, with three data points allocated for each type. 
The specific distribution of these data points is illustrated in Figure \ref{fig2b:PINN_domain}.
As outlined in Section \ref{sec:2.2_Assimilation}, the training of the PINN involves minimizing the discrepancy between the actual measured data at sensor locations and the output of the neural network at corresponding locations.
This discrepancy forms part of the loss function, enabling the assimilation of various types of measurement data to guide the training process of PINN.
Subsequently, the spatiotemporal coordinates of the target region serve as inputs for the trained PINN, enabling the reconstruction of the flow field within the designated target area.
Specifically, a uniform grid of $49 \times 25$ across the target area shown in Figure \ref{fig2b:PINN_domain} and 100 time steps spanning 0-100 seconds are used to assess the accuracy of the reconstructed unsteady flow fields.
The $RMSE$ is applied to evaluate the discrepancies in both velocity magnitude and velocity direction compared to the ground truth. 
Due to the inherent randomness in neural network training, each case is independently trained five times.
The $RMSE$ values presented in Figure \ref{fig3:Assi_MRSE} are the averages of these five training sessions, and the error bars indicate the standard deviation ($SD$).

The $RMSE$ quantifies the discrepancy between the reconstructed and actual flow fields, whereas the $SD$ reflects the variability of this error.
Statistical analysis presented in Figure \ref{fig3:Assi_MRSE} indicates that, overall, the accuracy of PINN reconstructions is generally commendable, albeit with discernible variations across different cases. 
In Case 1, where the training is exclusively based on LoS data, the results indicate lower accuracy and less consistency across experiments, in terms of both velocity magnitude and direction.
When additional types of measurement data are integrated with LoS data in Cases 2 to 4, there is a noticeable improvement in accuracy and consistency.
Among these, Case 2, which incorporates velocity vector data along with LoS data, stands out by achieving the highest accuracy and demonstrating superior consistency across multiple experiments.
This indicates that velocity vector data significantly enhances the PINN's ability to accurately reconstruct the flow field.
Case 4, which includes pressure data in addition to LoS data, shows a slight improvement in velocity magnitude accuracy compared to Case 3, which uses velocity component data.
However, Case 4 does not perform as well in terms of experimental consistency and accuracy in velocity direction as Case 3.
This suggests that while pressure data contributes to the accuracy of velocity magnitude, it may not be as effective in improving the accuracy of velocity direction, especially when compared to velocity component and velocity vector data.
Cases 5 to 7, which involve training with LoS data combined with two other types of measurement data, exhibit further improvements.
Within this group, Case 5, which incorporates both velocity vector and component data, shows the highest accuracy in terms of both velocity magnitude and direction.
While Case 6, which incorporates pressure data in addition to the measurements used in Case 2, results in only a slight improvement in the accuracy of velocity direction reconstruction, it also exhibits a marginal decrease in the accuracy of velocity magnitude reconstruction.
This also indicates that the efficiency of utilizing pressure data is relatively low.
In the governing equations \ref{eq:momentum_X} and \ref{eq:momentum_y}, only the pressure gradient term is present, meaning that the pressure is constrained through its gradient rather than being explicitly constrained.
Case 8, utilizing all four data types, demonstrates the utmost accuracy and consistency in repeated experiments compared to all other cases.
A comparative analysis of the statistical results from the eight cases reveals a trend of escalating accuracy in flow field reconstruction with the incremental inclusion of diverse measurement data types.
Moreover, with the addition of extra data, the standard deviation of multiple experiments also shows a trend of convergence, indicating that the inherent randomness of the neural network is weakening, and the consistency of repeated experiments has been improved.
These results underscore the value of incorporating a diverse range of measurement data in the training process of PINN for wind flow field reconstruction.
The ability to assimilate various types of data effectively enhances the performance of the proposed framework, leading to more accurate and consistent reconstructions.
This is particularly important in applications where precise flow field information is crucial, such as in wind energy resource assessment and environmental fluid dynamics.

\begin{figure*}[!htbp]
  \centering
  \subfloat[$t=30s$ ground truth]{
    \includegraphics[width=0.32\textwidth]{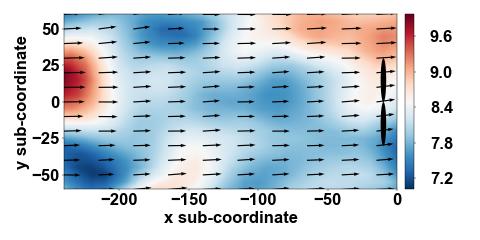} \label{fig:Assim_t30s_true} }
  \hfill
  \subfloat[$t=30s$ PINN ]{
    \includegraphics[width=0.32\textwidth]{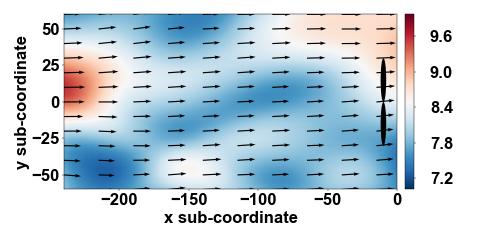} \label{fig:Assim_t30s_pred} }
  \hfill
  \subfloat[$t=30s$ point-wise error]{
    \includegraphics[width=0.32\textwidth]{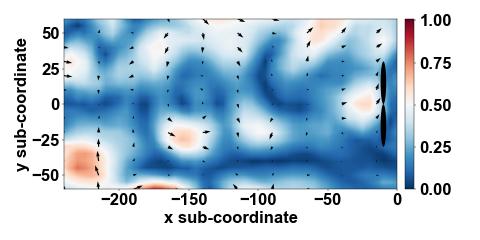} \label{fig:Assim_t30s_error} }

  \subfloat[$t=40s$ ground truth]{
    \includegraphics[width=0.32\textwidth]{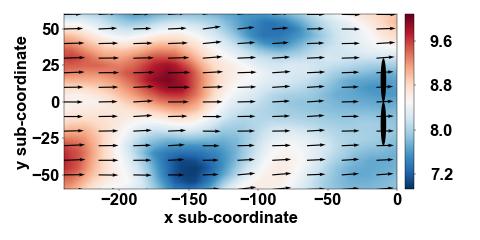} \label{fig:Assim_t40s_true} }
  \hfill
  \subfloat[$t=40s$ PINN ]{
    \includegraphics[width=0.32\textwidth]{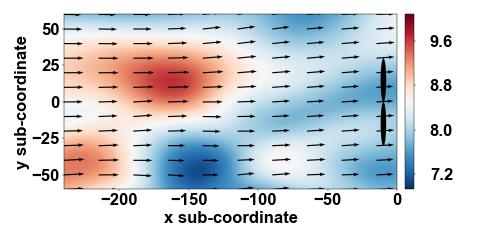} \label{fig:Assim_t40s_pred} }
  \hfill
  \subfloat[$t=40s$ point-wise error ]{
    \includegraphics[width=0.32\textwidth]{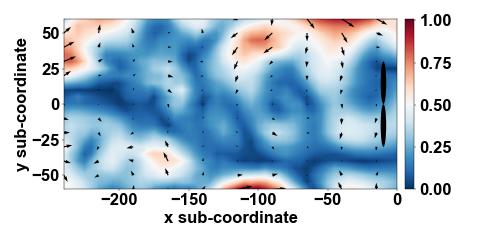} \label{fig:Assim_t40s_error} }

  \subfloat[$t=50s$ ground truth]{
    \includegraphics[width=0.32\textwidth]{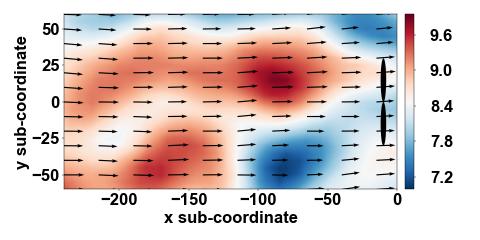} \label{fig:Assim_t50s_true} }
  \hfill
  \subfloat[$t=50s$ PINN]{
    \includegraphics[width=0.32\textwidth]{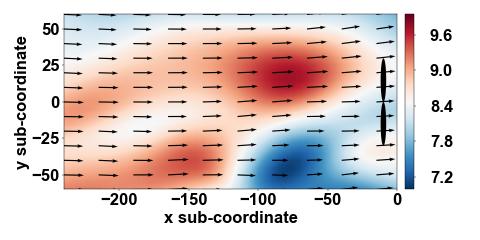} \label{fig:Assim_t50s_pred} }
  \hfill
  \subfloat[$t=50s$ point-wise error]{
    \includegraphics[width=0.32\textwidth]{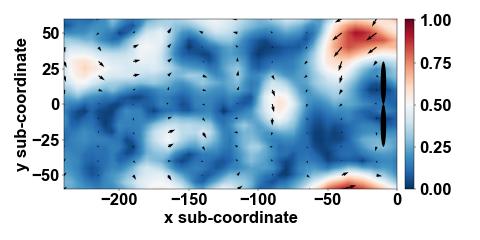} \label{fig:Assim_t50s_error} }

  \subfloat[$t=60s$ ground truth]{
    \includegraphics[width=0.32\textwidth]{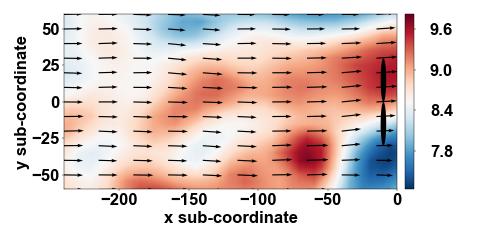} \label{fig:Assim_t60s_true} }
  \hfill
  \subfloat[$t=60s$ PINN]{
    \includegraphics[width=0.32\textwidth]{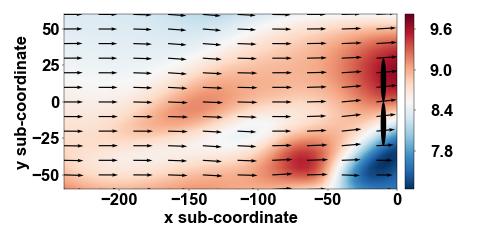} \label{fig:Assim_t60s_pred} }
  \hfill
  \subfloat[$t=60s$ point-wise error]{
    \includegraphics[width=0.32\textwidth]{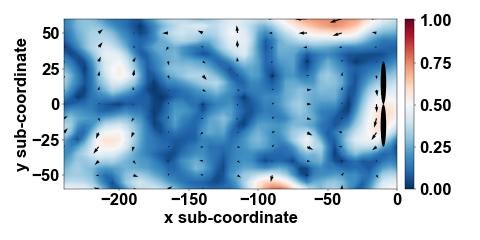} \label{fig:Assim_t60s_error} }
    
\caption{The comparison of the wind flow field reconstructed by the PINN trained in Case 8 with the ground truth. The values in the contour map represent the velocity magnitude $(m/s)$, while the arrows indicate the velocity direction.}
\label{fig4:Assim_flow_field}
\end{figure*}

Figure \ref{fig4:Assim_flow_field} presents the instantaneous flow field visualization for Case 8.
The first column of the figure shows the ground truth from simulations, while the second column depicts the flow field reconstructed by the trained PINN.
The third column illustrates the point-wise error.
The rows, from the first to the fourth, represent the instantaneous flow fields at times $t$ = 30 $s$, 40 $s$, 50 $s$, and 60 $s$, respectively.
Figure \ref{fig4:Assim_flow_field} indicates that each reconstructed snapshot exhibits a high degree of consistency with the corresponding actual snapshots.
During the training process of the neural network, various types of measurement data are assimilated, guided by the constraints of the parameterized N-S equations.
The network effectively learns the relationship between spatiotemporal coordinates and wind speed from sparse measurements, addressing the challenge of limited experimental measurements in capturing comprehensive flow field information.
It is particularly notable that in Figure \ref{fig:Assim_t30s_pred}, \ref{fig:Assim_t40s_pred}, \ref{fig:Assim_t50s_pred} and \ref{fig:Assim_t60s_pred}, the regions depicted in red, indicating high-speed flow structures, enter the target area at $t$ = 30 $s$ and progressively shift rightward, passing the turbine rotor at $t$ = 60 $s$.
This movement of the high-speed flow structure is accurately captured by the data assimilation framework, which is significant for both pre-construction wind turbine surveying and operational load control.
As reviewed in the literature \cite{meyers2022wind}, the observability of flow within wind farms is often limited by sensor capabilities and the availability of signals and channels.
Moreover, wind turbine control algorithms heavily rely on accurately perceiving the current flow field and turbine states.
Therefore, the proposed data assimilation framework, designed to optimize the use of limited and varied data types, holds significant potential in addressing this research challenge.

\begin{figure*}[!htbp]
  \centering
  \subfloat[$t=50s$ ground truth]{
    \includegraphics[width=0.32\textwidth]{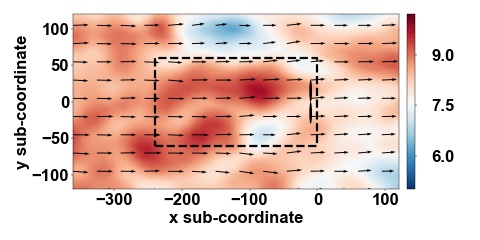} \label{fig:Fig_Ex_t049s_true} }
  \hfill
  \subfloat[$t=50s$ PINN]{
    \includegraphics[width=0.32\textwidth]{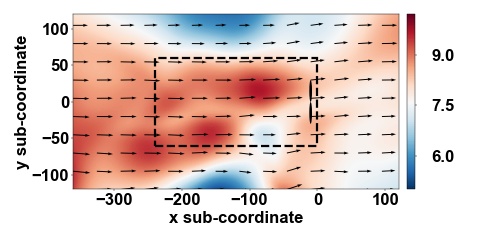} \label{fig:Fig_Ex_t049s_pred} }
  \hfill
  \subfloat[$t=50s$ point-wise error]{
    \includegraphics[width=0.32\textwidth]{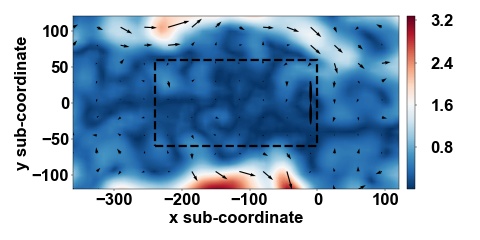} \label{fig:Fig_Ex_t049s_error} }
    
\caption{The comparison of the spatial extrapolated wind flow field by the PINN trained in Case 8 with the ground truth within the region $x \in [-6D,2D]$, $y \in [-2D,2D]$ at $t=50s$. The dotted rectangular frame represents the range of equation points. The values in the contour map represent the velocity magnitude $(m/s)$, while the arrows indicate the velocity direction.}
\label{fig:Assim_Ex}
\end{figure*}

The trained PINN demonstrates a remarkable capability to reconstruct flow fields not only within, but also extending beyond, the specific range of equation points utilized during its training phase.
This broadened applicative ability showcases the adaptability of the PINN model in handling extrapolative tasks in fluid dynamics.
As depicted in Figure \ref{fig:Assim_Ex}, we present a comparison between the reconstructed flow fields by the PINN and the ground truth within the extended region of $x \in [-6D,2D]$, $y \in [-2D,2D]$.
The black dotted line demarcates the boundary of the equation points, specifically set at $x \in [-4D,0]$ and $y \in [-D,D]$. 
It is noteworthy that outside the confines of the black dotted line, where no data points or specific equation points are provided during the training process, the PINN still managed to reconstruct the flow fields with a commendable degree of accuracy.
Moreover, the analysis of the point-wise errors reveals that larger discrepancies primarily occur in regions significantly distanced from the core range of equation points.
This observation suggests a limitation in the model's extrapolative performance, which tends to degrade as the distance from known data points increases.
However, despite these limitations, the results affirm that the proposed data assimilation framework effectively enhances the perception of the flow field dynamics.
This enhancement allows the network to infer wind speed distributions over an appreciably larger area than that covered by the direct measurements.

\begin{figure*}[!htbp]
  \centering
  \subfloat[$x=-210m$]{
    \includegraphics[width=0.32\textwidth]{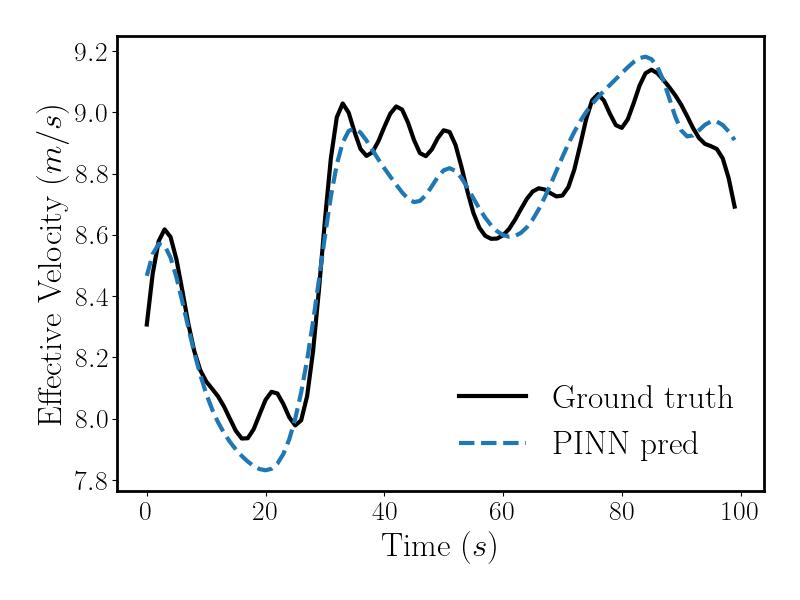} \label{fig:Assim_Ueff_x-210} }
  \hfill
  \subfloat[$x=-170m$]{
    \includegraphics[width=0.32\textwidth]{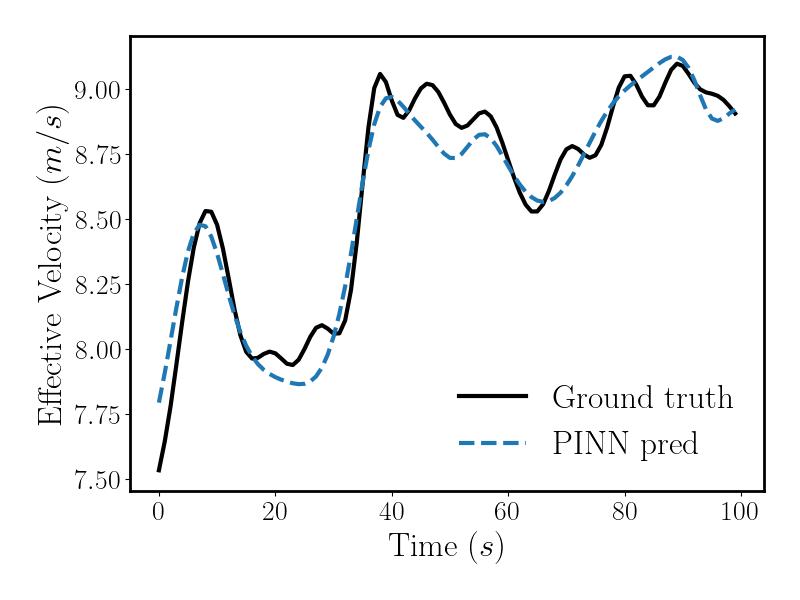} \label{fig:Assim_Ueff_x-170} }
  \hfill
  \subfloat[$x=-130m$]{
    \includegraphics[width=0.32\textwidth]{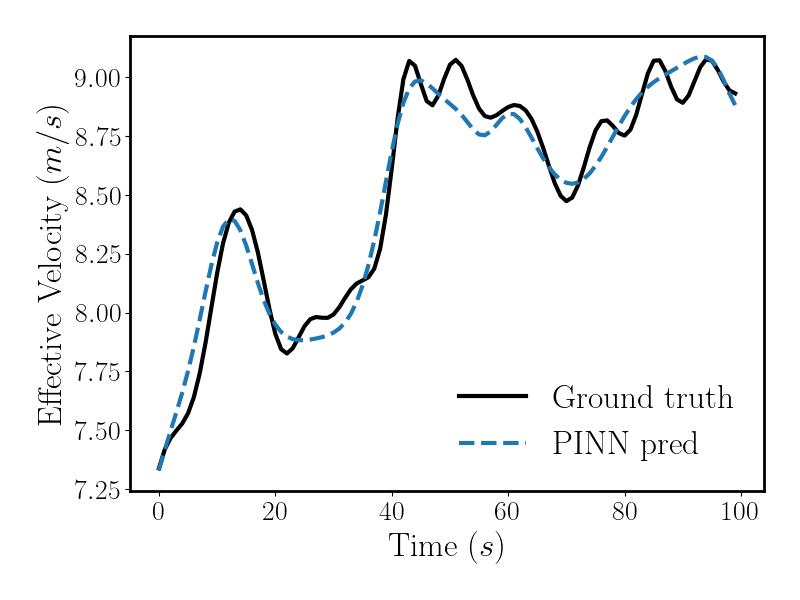} \label{fig:Assim_Ueff_x-130} }

  \subfloat[$x=-90m$]{
    \includegraphics[width=0.32\textwidth]{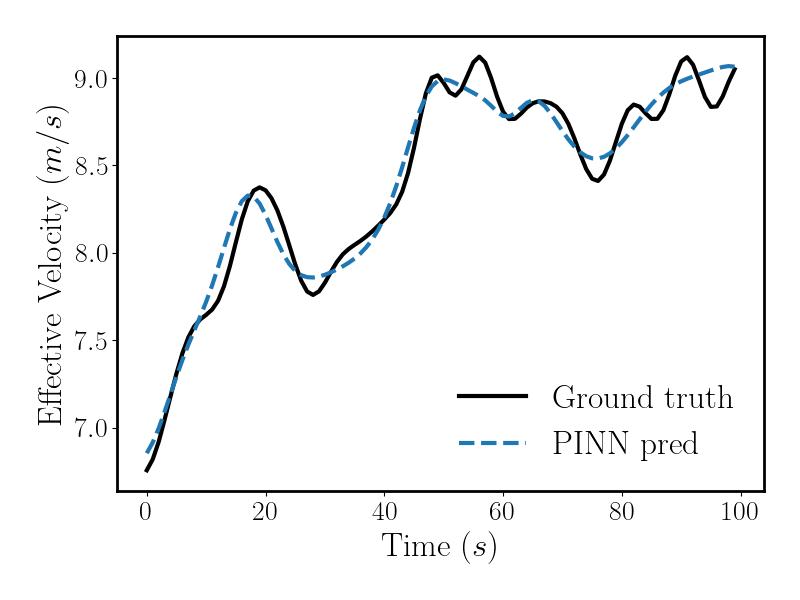} \label{fig:Assim_Ueff_x-90} }
  \hfill
  \subfloat[$x=-50m$]{
    \includegraphics[width=0.32\textwidth]{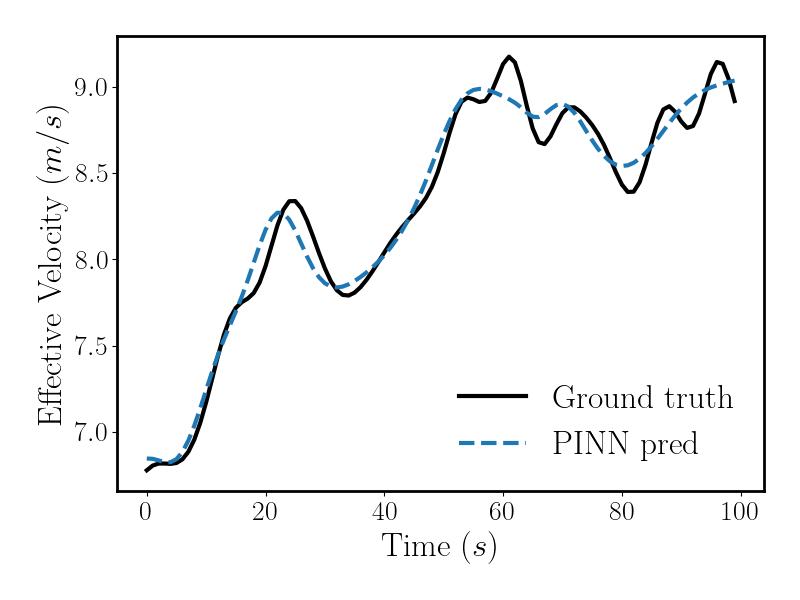} \label{fig:Assim_Ueff_x-50} }
  \hfill
  \subfloat[$x=-10m$]{
    \includegraphics[width=0.32\textwidth]{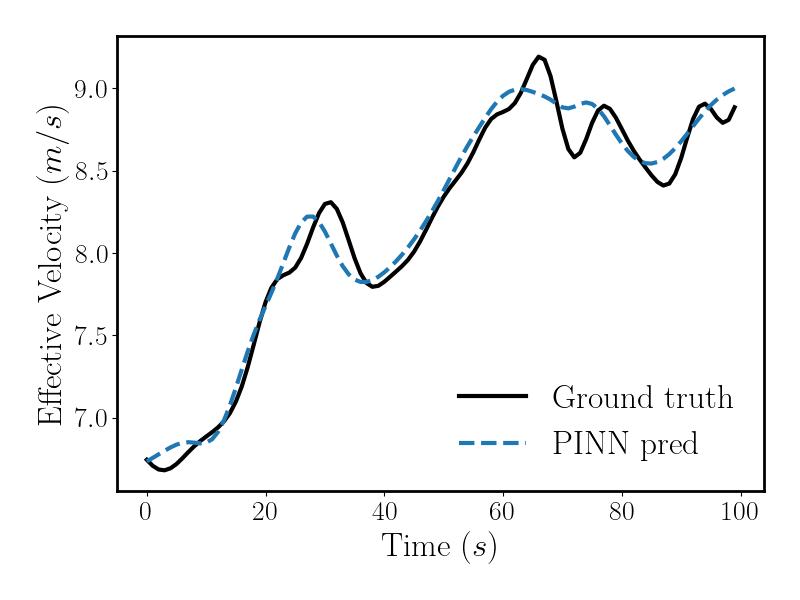} \label{fig:Assim_Ueff_x-10} }
    
\caption{The effective wind speed upstream of the wind turbine rotor. The predictions of PINN in Case 8 are compared with the ground truth simulated by SOWFA.}
\label{fig5:Assim_Ueff}
\end{figure*}

Given the capability of our framework to reconstruct a comprehensive flow field, an additional promising application emerges in deducing the effective wind speed.
This can be instrumental for power assessment or structural design purposes.
Effective wind speed is defined as the average of the wind speed projected onto the rotor plane:
\begin{equation}
{\overline u_{eff} (x,t)} = \frac{1}{{{N_y}}}\sum\limits_{i = 1}^{{N_y}} {u(x,{y_i},t)},\quad {y_i} \in \left( { - D/2,D/2} \right)
    \label{eq:Ueff}
\end{equation}
This equation calculates the average wind speed perpendicular to the rotor plane at a fixed $x$ coordinate and a specific moment in time.
Figure \ref{fig5:Assim_Ueff} compares the actual effective wind speeds at various $x$-coordinates with those predicted by the PINN in Case 8.
The $x$ coordinates of -210 $m$, -170 $m$, -130 $m$, -90 $m$, -50 $m$, and -10 $m$ represent positions 200 $m$, 160 $m$, 120 $m$, 80 $m$, 40 $m$, 0 $m$ upstream of the wind turbine rotor, respectively.
The effective wind speeds predicted by the PINN align closely with the actual values.
Specifically, at $x$ coordinates of -210 $m$, -170 $m$, -130 $m$, -90 $m$, -50 $m$, and -10 $m$, the maximum deviations in predicted $\overline u_{eff}$ are 3.1\%, 3.5\%, 3.0\%, 2.8\%, 2.7\%, and 3.6\%, occurring at 22 $s$, 2 $s$, 8 $s$, 14 $s$, 20 $s$, and 73 $s$, respectively.
The accuracy of the reconstructed flow field ensures the precision of the effective speed.
This highlights the benefit of reconstructing a detailed flow field, as it enables the extraction of specific flow field characteristics as needed.

\subsection{\label{sec:3.4_Influence of coef} The impact of loss function weights}
As indicated in Table \ref{tab:hyper}, the proposed data assimilation framework includes several hyperparameters that influence the accuracy of wind flow field reconstruction.
These hyperparameters can be finely tuned to suit specific applications.
Among these hyperparameters, the weights of the loss function, $\lambda_{pde}$ and $\lambda_{data}$, are unique to the PINN.
As per Equation (\ref{eq:loss}), $\lambda_{pde}$ and $\lambda_{data}$ serve as weights for equation loss $\mathcal{L}_{pde}$ and data loss $\mathcal{L}_{data}$, respectively. 
These two hyperparameters determine the relative contributions of $\mathcal{L}_{pde}$ and $\mathcal{L}_{data}$ to the overall loss function $\mathcal{L}$.
Essentially, they balance the optimization direction of the neural network, whether it leans more towards fitting the governing equations or aligning with the measurement data.
In Equation (\ref{eq:loss}), $\lambda_{pde}$ and $\lambda_{data}$ are explicitly written for clarity, but in practice, only their ratio $\lambda_{pde}/\lambda_{data}$ is required as an independent hyperparameter. 
In this section, $\lambda_{data}$ is fixed at 1, and $\lambda_{pde}$ is varied to adjust the weights.
The specific configurations are detailed in Table \ref{tab:CoefEq}.
The numerical experiments to examine the impact of loss function weights are based on Case 8, with $\lambda_{pde}$ ranging from 0.0001 to 10000, while all other hyperparameters remain consistent with those in Case 8.
\begin{table}[!htbp]
    \centering
    \caption{The numerical experiment setup for different loss function weights.}
    \renewcommand{\arraystretch}{1.5}
    \begin{tabular}{cccccccccc}
    \toprule
      & Case 8 & Case 8-1 & Case 8-2 & Case 8-3 & Case 8-4 & Case 8-5 & Case 8-6 & Case 8-7 & Case 8-8 \\ \midrule
     $\lambda_{pde}$ & 1 & 0.0001 & 0.001 & 0.01 & 0.1 & 10 & 100 & 1000 & 10000 \\
     $\lambda_{data}$ & 1 & 1 & 1 & 1 & 1 & 1 & 1 & 1 & 1\\ \bottomrule
    \end{tabular}
    \label{tab:CoefEq}
\end{table}

\begin{figure*}[!htbp]
\centering
  \subfloat[Wind speed magnitude]{
    \includegraphics[width=0.4\textwidth]{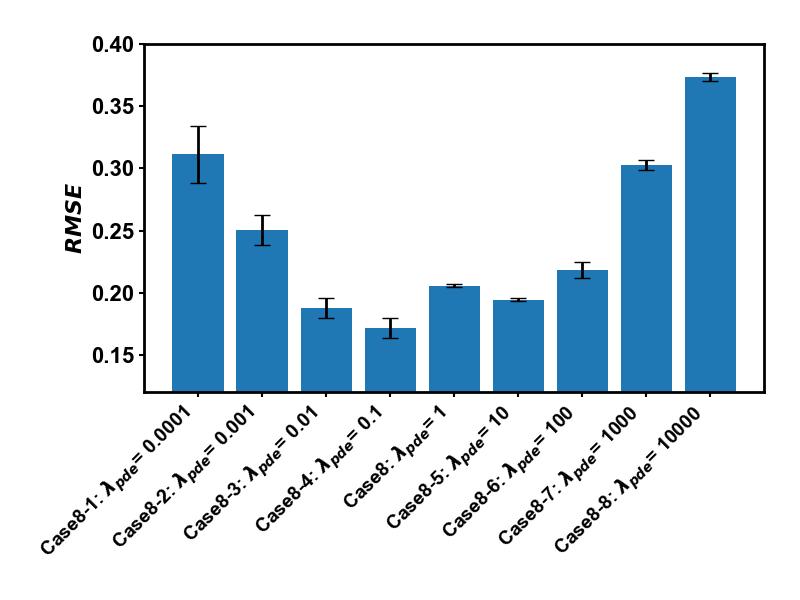} \label{fig3a:CoefEq_MRSE_mag} }
  \subfloat[Wind speed direction]{
    \includegraphics[width=0.4\textwidth]{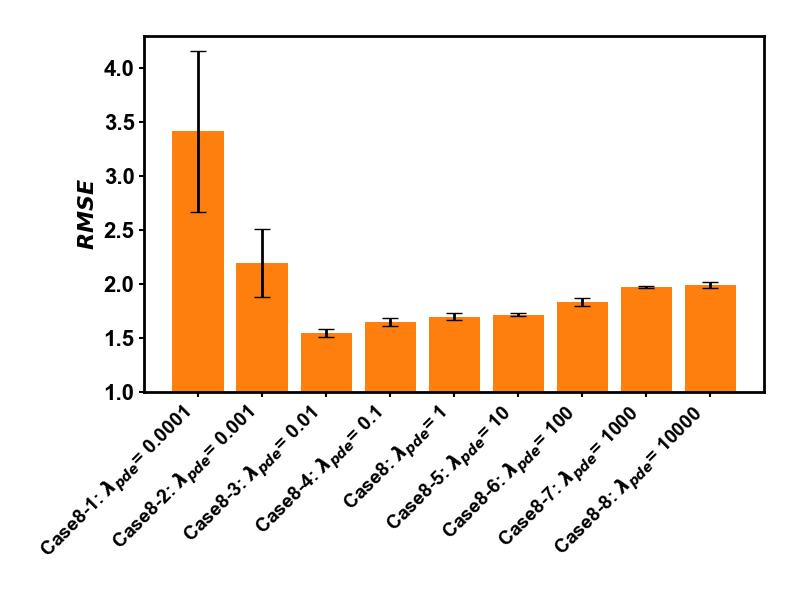} \label{fig3b:CoefEq_MRSE_dir} }
\caption{The influence of various loss function weights.}
\label{fig:CoefEq_MRSE}
\end{figure*}

Five independent training sessions are conducted for each case in this set of numerical experiments. 
Upon completion of the training, the reconstructed flow fields are compared with the ground truth, and the $RMSE$ and $SD$ of the reconstructed flow field errors are calculated. 
These results are presented in Figure \ref{fig:CoefEq_MRSE}.
The statistical analysis reveals that the weight of the loss function significantly influences the accuracy of the reconstruction.
By appropriately reducing the weight assigned to the equation loss $\mathcal{L}_{pde}$, a discernible improvement in the accuracy of the reconstructed flow field is achieved.
For instance, in Case 8-4 with $\lambda_{pde}$ = 0.1, the $RMSE$ for both velocity magnitude and direction decreases by 16.5\% and 3.2\%, respectively, compared to Case 8 with $\lambda_{pde}$ = 1. 
Similarly, in Case 8-3 with $\lambda_{pde}$ = 0.01, the $RMSE$ for velocity magnitude and direction decreases by 8.8\% and 9.1\%, respectively, compared to Case 8.
However, an excessive reduction in $\lambda_{pde}$ leads to a significant drop in the accuracy of the reconstructed flow field.
In addition, the consistency across multiple experiments deteriorates.
Furthermore, increasing $\lambda_{pde}$ also results in decreased accuracy of the reconstructed flow field, with the accuracy of velocity magnitude being more adversely affected than that of velocity direction.

\begin{figure*}[!htbp]
  \centering
  \subfloat[Case 8-2, $\lambda_{pde}=0.001$]{
    \includegraphics[width=0.32\textwidth]{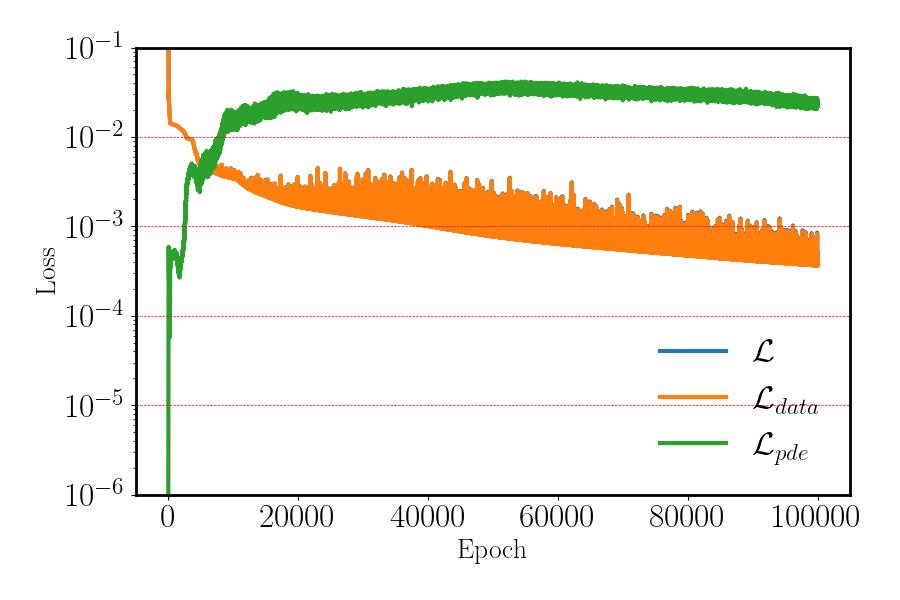} \label{fig:Loss_CoefEq_0.001} }
  \hfill
  \subfloat[Case 8-3, $\lambda_{pde}=0.01$]{
    \includegraphics[width=0.32\textwidth]{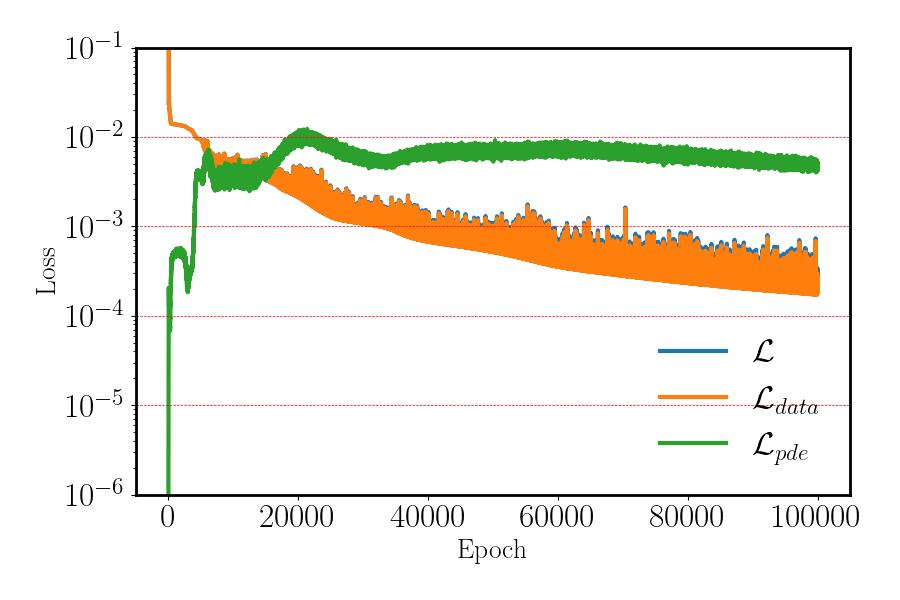} \label{fig:Loss_CoefEq_0.01} }
  \hfill
  \subfloat[Case 8-4, $\lambda_{pde}=0.1$]{
    \includegraphics[width=0.32\textwidth]{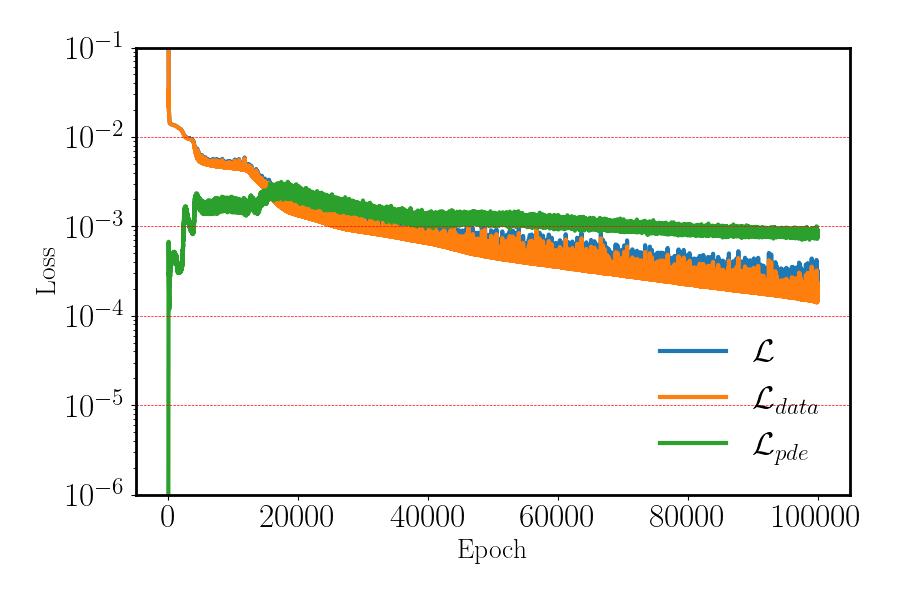} \label{fig:Loss_CoefEq_0.1} }

  \subfloat[Case 8, $\lambda_{pde}=1$]{
    \includegraphics[width=0.32\textwidth]{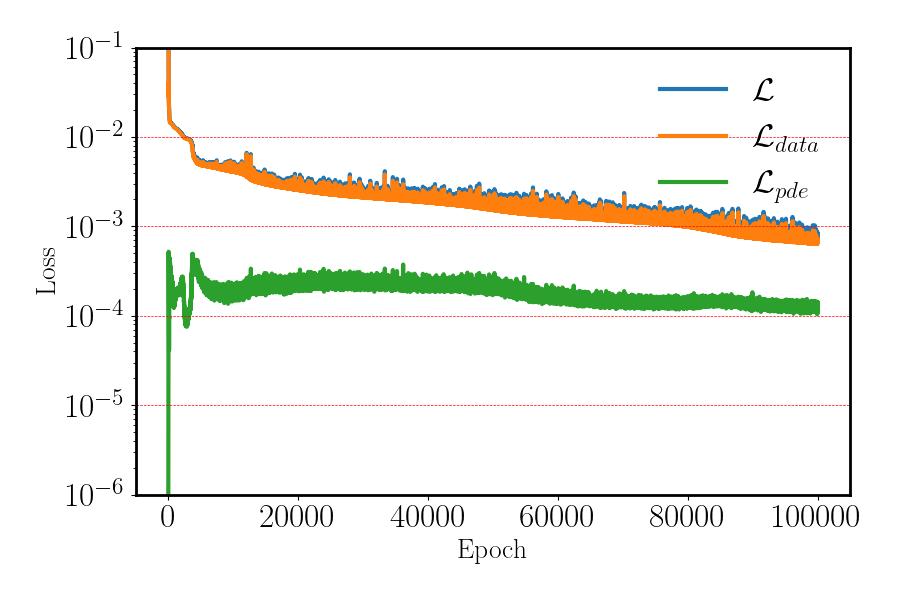} \label{fig:Loss_CoefEq_1} }
  \hfill
  \subfloat[Case 8-5, $\lambda_{pde}=10$]{
    \includegraphics[width=0.32\textwidth]{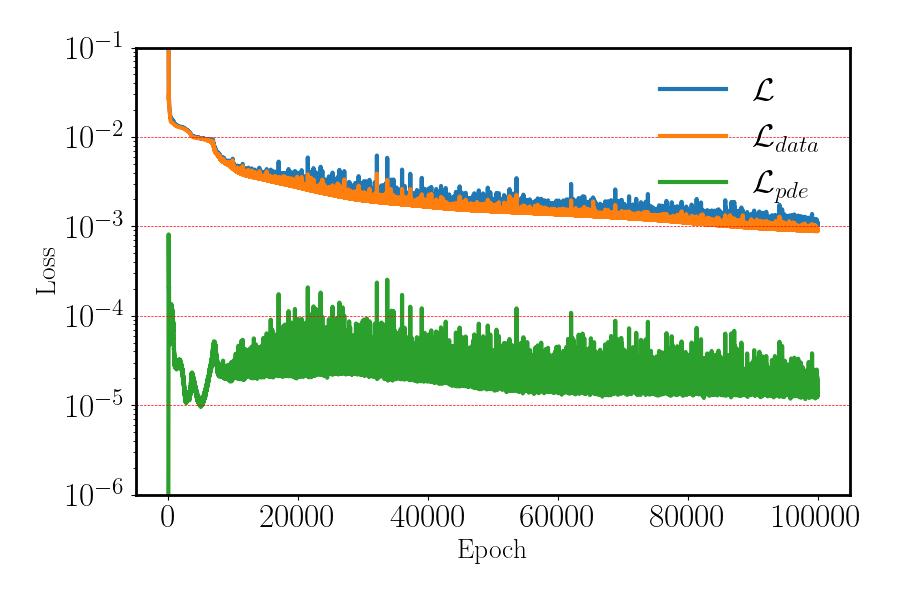} \label{fig:Loss_CoefEq_10} }
  \hfill
  \subfloat[Case 8-6, $\lambda_{pde}=100$]{
    \includegraphics[width=0.32\textwidth]{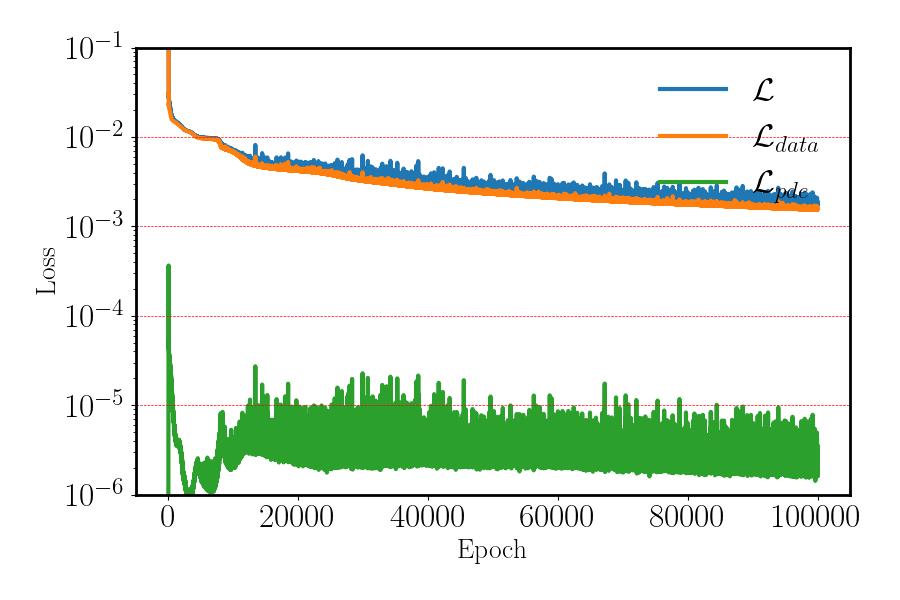} \label{fig:Loss_CoefEq_100} }
    
\caption{The changes in the loss function value during the training of Cases with $\lambda_{pde}$ ranging from 0.001 to 100.}
\label{fig:Loss_CoefEq}
\end{figure*}

To investigate the impact of loss function weights on the training process, Figure \ref{fig:Loss_CoefEq} displays the evolution of the loss function values during the training of cases with $\lambda_{pde}$ values ranging from 0.001 to 100.
Each case, as shown in Figure \ref{fig:Loss_CoefEq}, undergoes sufficient training, with the loss functions exhibiting convergence characteristics.
Figure \ref{fig:Loss_CoefEq_1} focuses on the training process of the model in Case 8, where both $\lambda_{pde}$ and $\lambda_{data}$ are set to 1.
In this scenario, the data loss stabilizes at approximately $10^{-3}$, while the equation loss converges near $10^{-4}$.
This pattern indicates a stronger tendency of the neural network's optimization process to adhere to equation constraints.
Consequently, a reduction in $\lambda_{pde}$ leads to a gradual shift in the neural network's optimization focus, moving from predominantly equation constraints towards placing more emphasis on the observed data, and vice versa.
For example, in Case 8-3, as depicted in Figure \ref{fig:Loss_CoefEq_0.01}, and Case 8-4, shown in Figure \ref{fig:Loss_CoefEq_0.1}, the equation constraints are relaxed, prompting the neural network to prioritize minimizing discrepancies with the observed data.
The flow fields reconstructed in Case 8-3 and Case 8-4 exhibit higher accuracy than Case 8, suggesting that these cases strike a closer balance between equation loss and data loss.
However, there is a threshold to how much $\lambda_{pde}$ can be reduced.
As illustrated in Figure \ref{fig:Loss_CoefEq_0.001}, when $\lambda_{pde}$ is decreased to 0.001, the balance is disrupted, leading to an excessive focus on minimizing data loss.
This imbalance results in a significant decrease in the accuracy of the reconstructed flow field.
Similarly, Case 8 demonstrates a marked deviation from the desired equilibrium, with a pronounced inclination towards equation loss.
Further increasing $\lambda_{pde}$ exacerbates this imbalance, leading to a more pronounced reduction in accuracy.

In summary, the proposed data assimilation framework is designed to simultaneously fit observational data and adhere to flow governing equations. 
The hyperparameter $\lambda_{pde}/\lambda_{data}$ plays a pivotal role in maintaining equilibrium between these two aspects during the optimization phase of the neural network.
This balance is crucial for optimally minimizing both $\mathcal{L}_{pde}$ and $\mathcal{L}_{data}$.
In practical scenarios, the ratio of $\lambda_{pde}$ to $\lambda_{data}$ should be adjusted based on the convergent values of $\mathcal{L}_{pde}$ and $\mathcal{L}_{data}$, which are ascertained through numerical experiments.
By fine-tuning this ratio to bring the convergent values of $\mathcal{L}_{pde}$ and $\mathcal{L}_{data}$ closer together, the accuracy of the model can be further enhanced.
As a result, this approach leads to a harmoniously balanced data assimilation model that not only corresponds closely with measured data but also physically adheres to the constraints of governing equations.

\subsection{\label{sec:3.5_transfer_learning} Assimilating future measurements by transfer learning}
Theoretically, the structure of the PINN enables the direct prediction of future flow field data when a future time point is inputted.
However, this method is only effective over a very short period.
As time progresses, the accuracy of the predictions rapidly decreases.
Zhang and Zhao \cite{zhang2021spatiotemporal} has attempted to directly use a trained PINN for short-term prediction of wind flow fields.
The time range of the training set was from 0 $s$ to 100 $s$, and the forecasting results indicated significant deviations from actual values after 115 $s$.
Our previous research \cite{yan2022inferring} also reveals the weak predictive capability of PINN.
To overcome this limitation, the proposed data assimilation framework incorporates transfer learning.
This approach allows a pre-trained model to be further trained with real-time measurements in an online mode.
The objective of this strategy is to achieve online reconstruction to improve long-term accuracy while minimizing the number of training steps required.
This makes it feasible for deployment in an online system, offering a more dynamic and responsive approach to wind flow field reconstruction.

In this section, the most accurate model derived from Case 8-4 is selected as the pre-trained model for transfer learning training. 
Additional flow field data spanning 100 $s$ is generated via the SOWFA simulator, with measurement data collected in the same manner as in Case 8-4.
Considering the scenario of online deployment, the training duration for transfer learning is deliberately kept shorter than the physical flow time.
As a result, the training is limited to 1000 steps, ensuring that the time consumption is controlled to be within 100 $s$.
Transfer learning often employs a parameter-freezing strategy \cite{vrbanvcivc2020transfer}.
In this approach, certain parameters of the pre-trained model are fixed, while the remains are fine-tuned.
This technique accelerates the training process by reducing the number of parameters that need to be trained.
However, this increased efficiency might come at the cost of some accuracy.
The proposed data assimilation framework also allows for the configuration of frozen layers.
Considering that the pre-trained model comprises 10 hidden layers, the number of frozen layers $N_f$ in the numerical experiments varies, with configurations of 0, 2, 4, 6, and 8 layers.
In these settings, the first $N_f$ hidden layers of the pre-trained model are frozen, and the remaining layers undergo training.

\begin{figure*}[!htbp]
\centering
  \subfloat[$RMSE$ of Wind speed magnitude]{
    \includegraphics[width=0.32\textwidth]{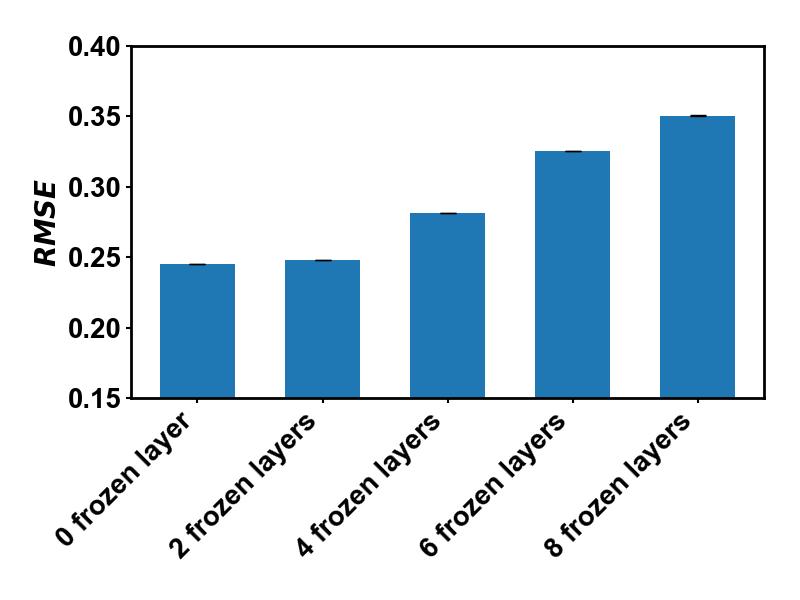} \label{fig3a:Trans_MRSE_mag} }
  \hfill
  \subfloat[$RMSE$ of Wind speed direction]{
    \includegraphics[width=0.32\textwidth]{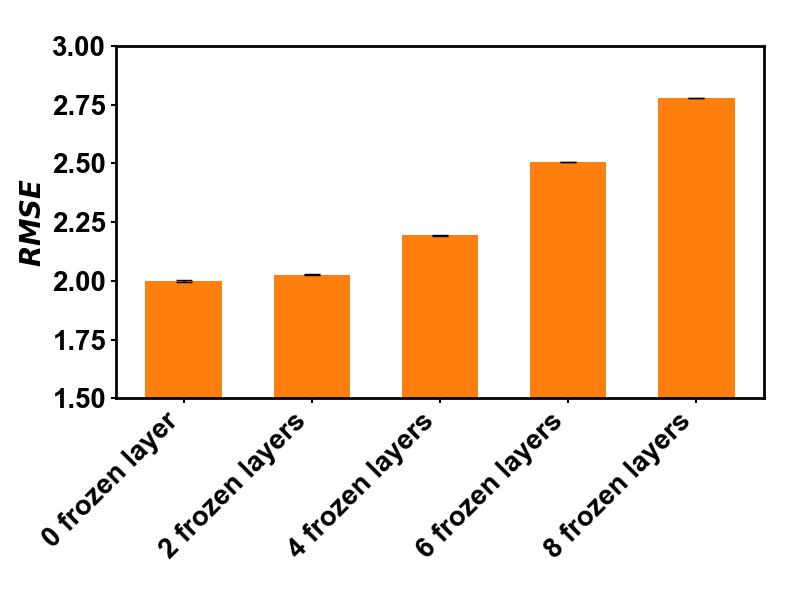} \label{fig3b:Trans_MRSE_dir} }
  \hfill
  \subfloat[Time consumption for transfer learning]{
    \includegraphics[width=0.32\textwidth]{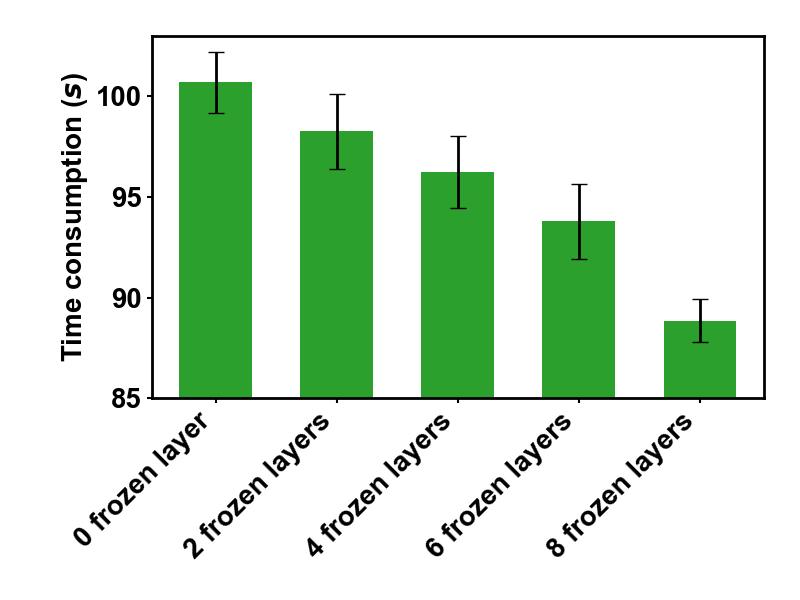} \label{fig3b:Trans_time} }
\caption{The $RMSE$ and time consumption of transfer learning. The $RMSE$ is compared the reconstructed wind flow fields by the transferred PINN and the ground truth, measured from 100 $s$ to 200 $s$. Time consumption is statistics of 1000 transfer learning steps.}
\label{fig:Trans_MRSE}
\end{figure*}

The statistical outcomes of the transfer learning experiments are presented in Figure \ref{fig:Trans_MRSE}.
The $RMSE$ quantifies the difference between the reconstructed flow fields and the ground truth over the time interval from 100 $s$ to 200 $s$.
Figures \ref{fig3a:Trans_MRSE_mag} and \ref{fig3b:Trans_MRSE_dir} demonstrate that the flow field reconstructed by the PINN after transfer learning closely is close to the actual flow field.
However, it is observed that increasing the number of frozen layers results in a noticeable decline in accuracy, without significant time savings.
The absence of substantial time savings is attributed to the necessity for the PINN to compute equation residuals.
This process, particularly the calculation of second-order derivatives during backpropagation, is inherently time-consuming.
Therefore, it is recommended to limit the freezing of layers to no more than two.
All transfer learning processes are still carried out on the NVIDIA Tesla V100 GPU.
The training of the transferred PINN with no frozen layers for 1000 steps takes approximately 100 $s$.
This duration could potentially be reduced with more powerful computing resources.
In contrast, the time required to reconstruct the flow field for 100 $s$ is about 0.2 $s$, which is negligible compared to the training duration.

\begin{figure*}[!htbp]
  \centering
  \subfloat[$x=-210m$]{
    \includegraphics[width=0.32\textwidth]{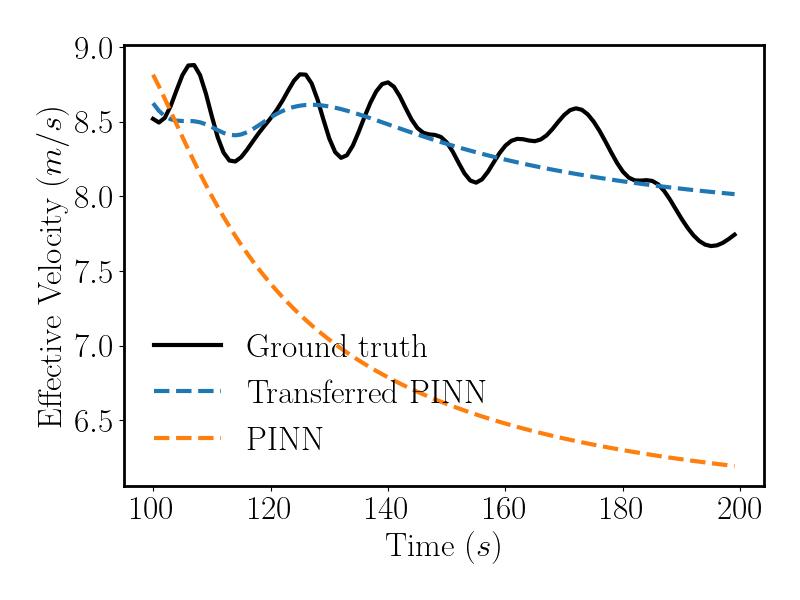} \label{fig:transF0_U_eff_x-210} }
  \hfill
  \subfloat[$x=-170m$]{
    \includegraphics[width=0.32\textwidth]{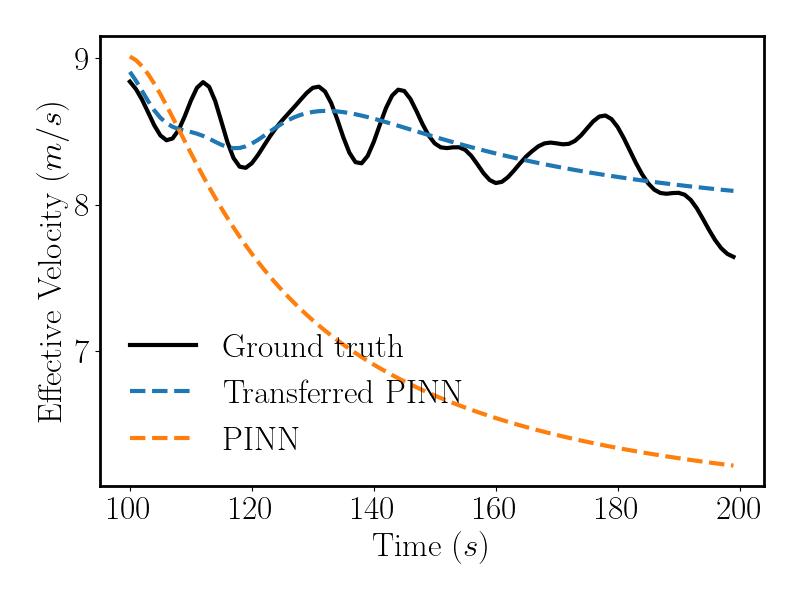} \label{fig:transF0_U_eff_x-170} }
  \hfill
  \subfloat[$x=-130m$]{
    \includegraphics[width=0.32\textwidth]{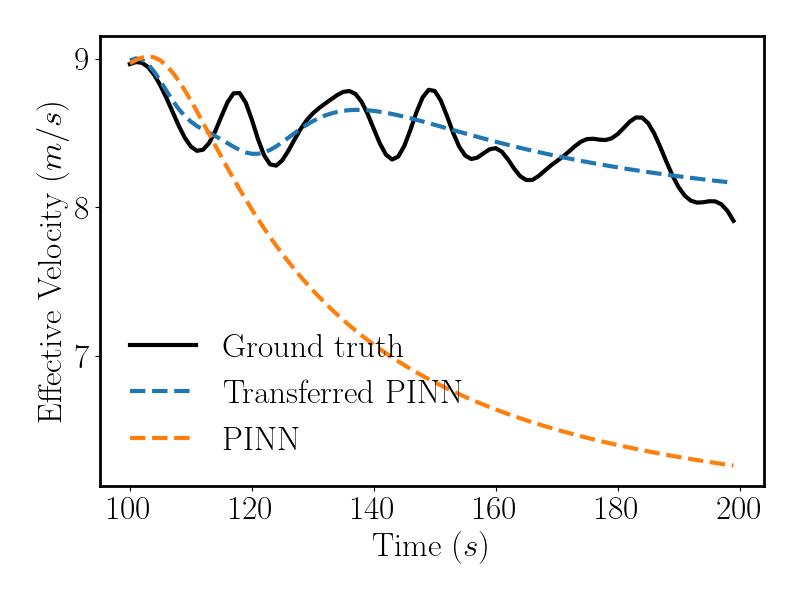} \label{fig:transF0_U_eff_x-130} }

  \subfloat[$x=-90m$]{
    \includegraphics[width=0.32\textwidth]{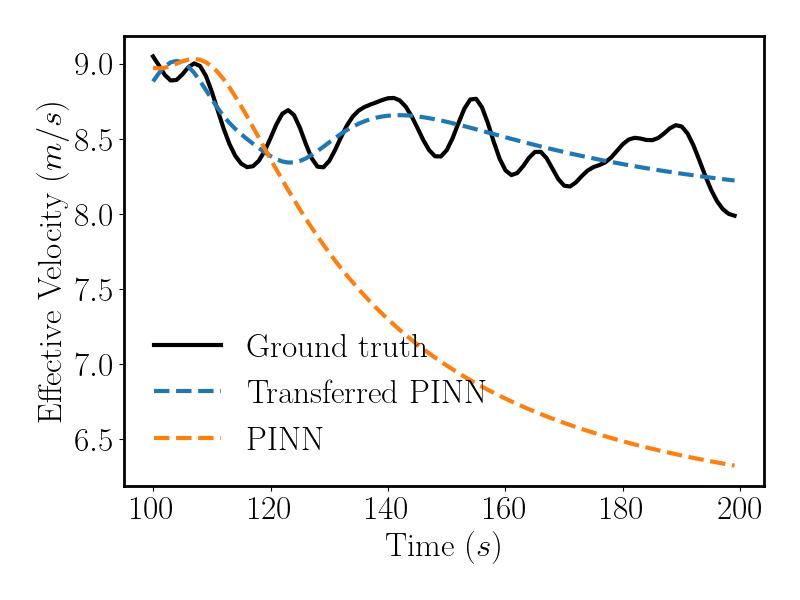} \label{fig:transF0_U_eff_x-90} }
  \hfill
  \subfloat[$x=-50m$]{
    \includegraphics[width=0.32\textwidth]{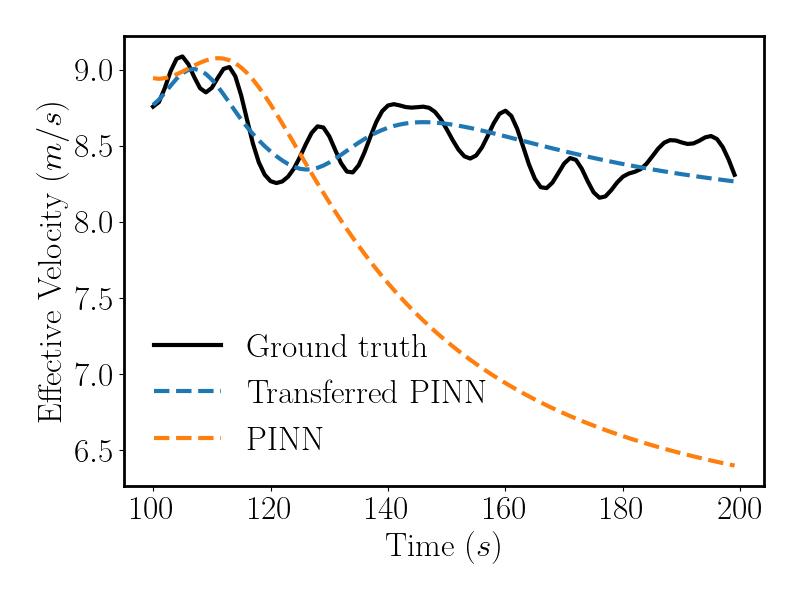} \label{fig:transF0_U_eff_x-50} }
  \hfill
  \subfloat[$x=-10m$]{
    \includegraphics[width=0.32\textwidth]{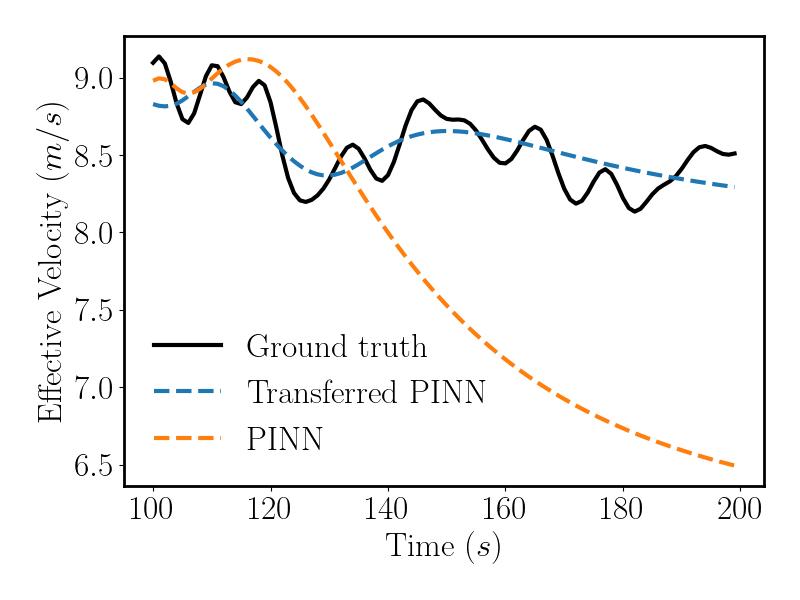} \label{fig:transF0_U_eff_x-10} }
\caption{The effective wind speed upstream of the wind turbine rotor from 100 $s$ to 200 $s$. The predictions of PINN and transferred PINN with no frozen layers are compared with the ground truth.}
\label{fig:Trans_Ueff}
\end{figure*}

\begin{figure*}[!htbp]
  \centering
  \subfloat[$t=120s$ ground truth]{
    \includegraphics[width=0.32\textwidth]{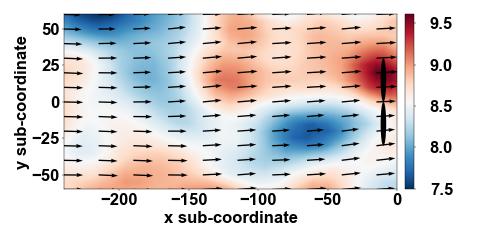} \label{fig:trans_t20s_true} }
  \hfill
  \subfloat[$t=120s$ transferred PINN]{
    \includegraphics[width=0.32\textwidth]{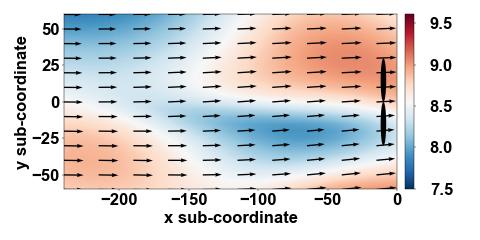} \label{fig:trans_t20s_pred} }
  \hfill
  \subfloat[$t=120s$ PINN ]{
    \includegraphics[width=0.32\textwidth]{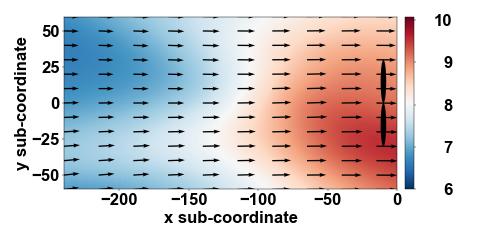} \label{fig:nt_t20s_pred} }
    
  \subfloat[$t=130s$ ground truth]{
    \includegraphics[width=0.32\textwidth]{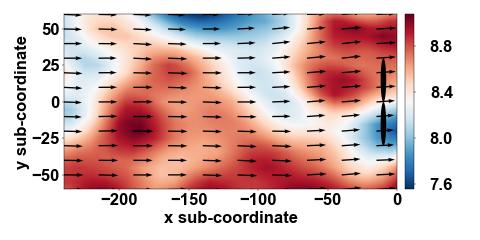} \label{fig:trans_t30s_true} }
  \hfill
  \subfloat[$t=130s$ transferred PINN]{
    \includegraphics[width=0.32\textwidth]{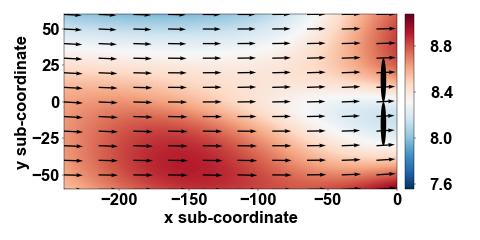} \label{fig:trans_t30s_pred} }
  \hfill
  \subfloat[$t=130s$ PINN ]{
    \includegraphics[width=0.32\textwidth]{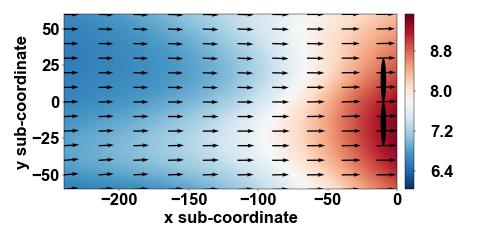} \label{fig:nt_t30s_pred} }
\caption{The comparison of wind flow field between the ground truth and reconstructed by the PINN with or without transfer learning. The values in the contour map represent the velocity magnitude $(m/s)$, while the arrows indicate the velocity direction.}
\label{fig:trans_flow_field}
\end{figure*}

To evaluate the accuracy of the transferred PINN in predicting effective wind speeds, the transferred PINN with no frozen layers is employed to reconstruct the flow field from 100 $s$ to 200 $s$ and subsequently calculate the effective wind speeds at a series of locations.
These results are compared with direct predictions from the PINN and actual results from SOWFA, as depicted in Figure \ref{fig:Trans_Ueff}.
The findings reveal that the accuracy of direct predictions from the PINN rapidly drops over time, significantly underestimating the effective wind speed.
In contrast, the PINN after transfer learning demonstrates a heightened ability to predict effective wind speeds with better accuracy.
The trends of predicted effective wind speed show good consistency with the actual values simulated by SOWFA.
At positions x=-210 $m$, -170 $m$, -130 $m$, -90 $m$, -50 $m$, and -10 $m$, the maximum errors of predicted $u_{eff}$ compared to the actual values are 5.1\%, 5.9\%, 4.4\%, 4.0\%, 3.3\%, and 3.7\%, respectively, occurring at 73 $s$, 100 $s$, 19 $s$, 24 $s$, 67 $s$, and 73 $s$.
However, it is noted that the prediction of velocity fluctuations is not as precise as the fully trained PINN in Sections \ref{sec:3.2_SinVar}, \ref{sec:3.3_Performance of data assimilation} and \ref{sec:3.4_Influence of coef}. 
The reconstructed instantaneous flow fields, as displayed in Figure \ref{fig:trans_flow_field}, corroborate this observation.
While the wind speed distribution and direction reconstructed by the PINN with transfer learning generally resemble the ground truth, there are noticeable discrepancies, particularly in areas of high-speed flow.
The reconstructed wind flow field tends to exhibit a degree of smoothness, lacking the intricate details present in the actual flow field.
In contrast, the wind speed distribution reconstructed by the PINN model without transfer learning shows more significant deviations from the ground truth.
The results have demonstrated the capability of the PINN, after transfer learning, to reconstruct the evolution of the flow field over a longer period.
In the context of online deployment, where the network cannot be fully trained within a time frame shorter than the physical flow duration, transfer learning emerges as a promising solution.
It has the potential to address the long-term prediction limitations inherent in PINN, offering a viable approach for online reconstruction in practical applications.

\section{Conclusion}
The data assimilation framework proposed in this paper provides a comprehensive approach for assimilating various types of experimental measurement data in wind energy applications.
The core structure of the framework, PINN, is trained under the dual constraints of measurement data and flow governing equations.
It supports four common types of measurement data: LoS wind speed, velocity vector, velocity component, and pressure.
The parameterized N-S equations are employed as the flow governing equations, with the artificial viscosity directly inferred by the network.
The training objective is to minimize both the discrepancy between network outputs and measurement data, as well as the residuals of the flow governing equations.
The trained network is capable of reconstructing a complete flow field and deriving any expected flow field information, such as the effective wind speed at specific locations.
The reconstructed flow field exhibits strong consistency with the flow field simulated using SOWFA.
Consequently, the effective wind speeds at different locations also display a high level of agreement.
Based on the current sensor configuration, as the variety of available measurement data increases, there is a corresponding improvement in the accuracy of the reconstructed flow field.
The addition of velocity vector data notably enhances the accuracy of the flow field reconstruction more than any other type of data.
This is followed by the inclusion of velocity component data, which also positively impacts accuracy, albeit to a lesser extent.
Due to the fundamental nature of momentum equations, only pressure gradients are included, not the pressure itself.
Consequently, the contribution of measured pressure data to improving accuracy is less significant compared to velocity data.

The influence of the distinctive hyperparameters of PINN, $\lambda_{pde}$ and $\lambda_{data}$, is investigated.
These two weights represent the weight coefficients for equation loss and data loss, respectively.
The ratio of the weights of equation loss to data loss, $\lambda_{pde}/\lambda_{data}$, varies from 0.0001 to 10000.
The results indicate that this ratio determines whether the network's optimization direction is more inclined towards data or loss.
It is beneficial for the accuracy of flow field reconstruction to balance the equation loss and data loss as much as possible.
An excessive bias towards either aspect inevitably leads to a decrease in accuracy.
To enhance accuracy in practical applications, fine-tuning the ratio $\lambda_{pde}/\lambda_{data}$ based on the convergent values of data loss and equation loss is recommended.

Notably, the integration of transfer learning enables PINN to reconstruct flow fields over extended periods, marking a shift from exclusively offline to potential online deployment.
The pre-trained PINN is deployed online and trained on the dataset of real-time measurement data over a certain period.
The transfer learning duration required is less than the actual physical flow time, and the model achieves acceptable accuracy in reconstructing the flow field for this period.
At the wind turbine site, the maximum error between the effective wind speed predicted online and the actual wind speed is only 3.7\%.
This represents a significant improvement compared to models that have not undergone transfer learning.
Therefore, the proposed framework demonstrates its potential for online deployment, making it a viable component of wind turbine online control systems.

Owing to the versatility of this framework, it can be adapted to systems governed by different PDEs, such as those involving wave, tidal, and other forms of energy, simply by altering the corresponding equations and measurement methods.
To further promote the utilization of renewable energy, the framework has been fully open-sourced, and collaborative development is encouraged.

\section*{CRediT authorship contribution statement}
\textbf{Chang Yan}: Conceptualization, Methodology, Investigation, Formal analysis, Software, Validation, Visualization, Writing - original draft.
\textbf{Shengfeng Xu}: Conceptualization, Methodology, Formal analysis, Software, Writing - original draft.
\textbf{Zhenxu Sun}: Funding acquisition, Methodology, Project administration, Supervision, Writing - review \& editing.
\textbf{Thorsten Lutz}:  Methodology, Investigation, Formal analysis, Supervision, Writing - review \& editing.
\textbf{Dilong Guo}: Funding acquisition, Validation, Visualization, Writing - review \& editing.
\textbf{Guowei Yang}: Funding acquisition, Investigation, Supervision, Writing - review \& editing.

\section*{Declaration of competing interest}
The authors declare that they have no known competing financial interests or personal relationships that could have appeared to influence the work reported in this paper.

\section*{Acknowledgments}
This research received funding from the National Key Research and Development Program of China under Grant No. 2022YFB2603400.
The first author is grateful for the support provided by the Chinese Academy of Sciences (CAS) and the German Academic Exchange Service (DAAD) Joint Fellowship, Grant No. 91842771, which facilitated research activities in Germany.
The authors extend special thanks to Dr. Jincheng Zhang from the University of Warwick for providing the simulation cases of wind flow fields, which greatly contributed to this research.
The authors also wish to express their appreciation to Prof. Dr.-Ing. Ewald Krämer from the University of Stuttgart for his insightful review of the manuscript and valuable feedback.

\appendix   

\section*{Appendix A. The robustness to turbine rotor diameters $D$}
As described in Section 3.1, the aerodynamics of the wind turbine rotor are not considered in this paper.
Therefore, the reference length for non-dimensionalization, the rotor diameter $D=60 m$, corresponds to the dimensions typically associated with medium-sized wind turbines.
This section examines the impact of varying rotor diameters on the accuracy of flow field reconstruction.

\begin{figure*}[!htbp]
\centering
  \subfloat[$RMSE$ of Wind speed magnitude]{
    \includegraphics[width=0.4\textwidth]{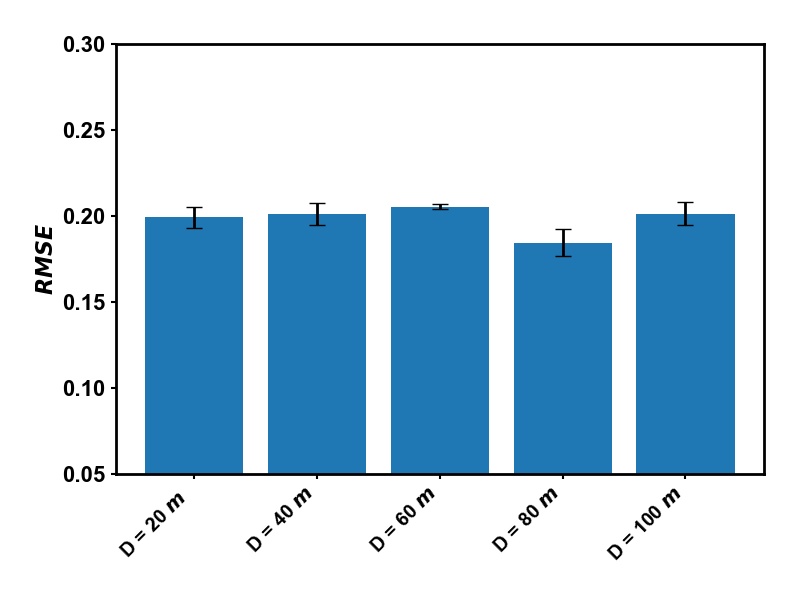} \label{figAa:Dim_MRSE_mag} }
  \subfloat[$RMSE$ of Wind speed direction]{
    \includegraphics[width=0.4\textwidth]{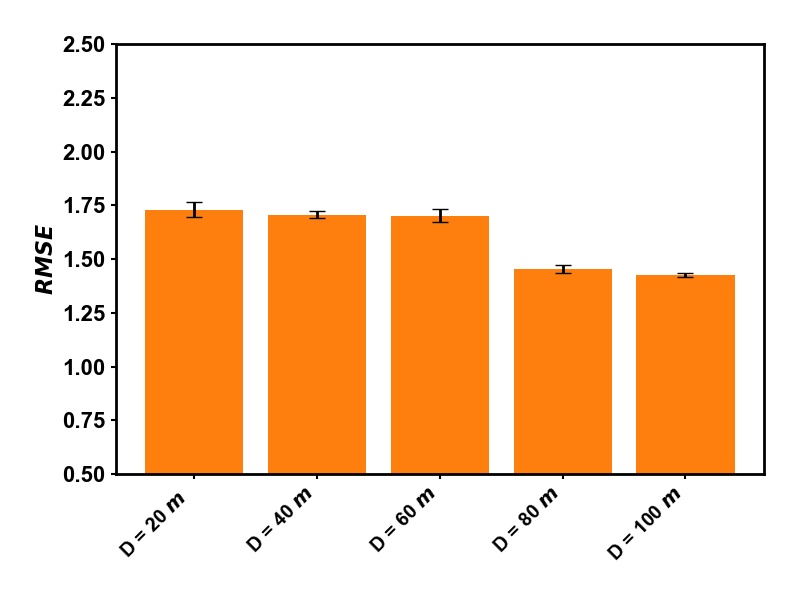} \label{figAb:Dim_MRSE_dir} }
\caption{The $RMSE$ under varying reference turbine rotor diameters $D$. The $RMSE$ values represent the relative root mean square errors between the reconstructed wind flow fields generated by the trained PINN and the ground truth. These $RMSE$ values are computed as the averages of five independent training sessions, with error bars indicating the standard deviations across these sessions.}
\label{figA:Dim_MRSE}
\end{figure*}

Rotor diameters of 20 $m$, 40 $m$, 60 $m$, 80 $m$, and 100 $m$ are evaluated, with the selection of data points and network training settings consistent with those employed in Case 8.
Each case with various rotor diameter is subject to five independent training sessions.
Figure \ref{figA:Dim_MRSE} presents the $RMSE$ of the reconstructed flow fields.
The results indicate that the $RMSE$ for other reference lengths closely aligns with that of Case 8, where $D=60 m$, and does not exceed the $RMSE$ observed in Case 8.
This suggests that the proposed data assimilation framework is robust with respect to the reference length $D$.

\section*{Appendix B. Reconstructed pressure and inferred artificial viscosity}
The primary objective of this study is to assimilate sparse and diverse measurements to wind speed fields. 
However, as depicted in Figure \ref{fig1:framework}, the data assimilation structure also enables the extraction of pressure fields $p$ and the estimation of artificial viscosity coefficients $\nu_{\eta}$, which are outputs of the network.
This appendix discusses the reconstructed pressure fields and the inferred artificial viscosity coefficients.

\begin{figure*}[!htbp]
\centering
  \subfloat[$t=60s$ ground truth]{
    \includegraphics[width=0.32\textwidth]{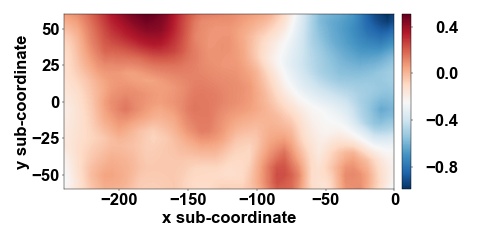} \label{figBa:Fig_p_true_t60s} }
  \hfill
  \subfloat[$t=60s$ PINN]{
    \includegraphics[width=0.32\textwidth]{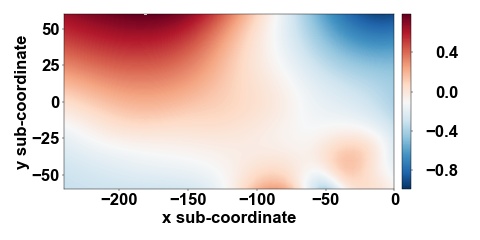} \label{figBb:Fig_p_pred_t60s} }
  \hfill
  \subfloat[$t=60s$ point-wise error]{
    \includegraphics[width=0.32\textwidth]{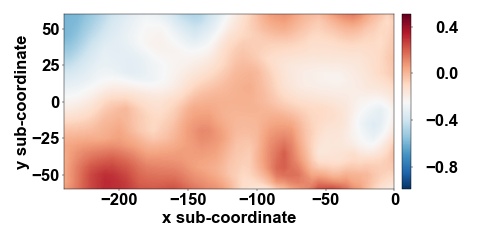} \label{figBc:Fig_p_error_t60s} }
\caption{The comparison of the pressure field reconstructed by the PINN trained in Case 8-4 with the ground truth. The values in the contour map represent the dimensional pressure ($Pa$).}
\label{figB:Fig_p_t60s}
\end{figure*}

Figure \ref{figB:Fig_p_t60s} illustrates the reconstructed pressure fields. 
The results generally indicate a satisfactory reconstruction of the pressure fields, yet notable errors are evident.
On one hand, the availability of pressure data is limited to only three measurement points, which restricts the extent of information that can be learned.
On the other hand, within the flow governing equations, only the gradient of pressure with respect to spatial coordinates is utilized.
This represents a weaker constraint compared to velocity, which further challenges the accuracy of pressure field reconstructions.

\begin{figure*}[!htbp]
\centering
  \subfloat[turbulent viscosity computed by SOWFA]{
    \includegraphics[width=0.48\textwidth]{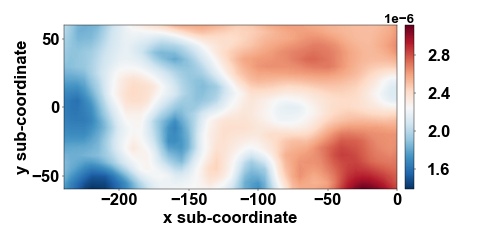} \label{figBa:Fig_nu_sgs_t60s} }
  \hfill
  \subfloat[artificial viscosity inferred by PINN]{
    \includegraphics[width=0.48\textwidth]{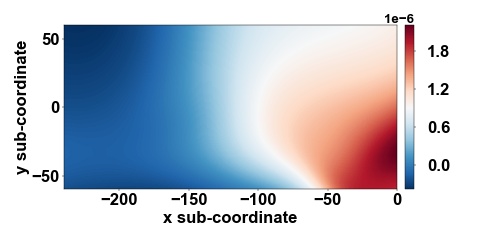} \label{figBb:Fig_nu_infer_t60s} }
\caption{The comparison of the artificial viscosity inferred by the PINN trained in Case 8-4 with the turbulent viscosity computed by SOWFA at $t = 60 s$. The values in the contour map represent the dimensional viscosity ($m^2/s$).}
\label{figB:Fig_nu_t60s}
\end{figure*}

Figure \ref{figB:Fig_nu_t60s} displays the artificial viscosity $\nu_{\eta}$ distribution inferred by the trained PINN in Case 8-4, alongside the turbulent viscosity distribution $\nu_{sgs}$ computed using LES in the SOWFA.
Both distributions do not represent actual physical viscosity but serve as control parameters that are instrumental in modeling small-scale turbulence.

The comparison depicted in Figure \ref{figB:Fig_nu_t60s} reveals that the trends in the distribution of artificial viscosity closely align with those of turbulence viscosity, both indicative of viscous dissipation effects.
However, the numerical values of the artificial viscosity inferred by the PINN are generally lower than those derived from LES-based turbulence models.
This suggests that the PINN, which utilizes a parameterized Navier-Stokes equation as a physical constraint, tends to employ lower dissipation effects compared to traditional turbulence models.

\section*{Appendix C. Examples of further fine-tuning data weights}
\begin{figure*}[!htbp]
\centering
  \subfloat[$RMSE$ of Wind speed magnitude]{
    \includegraphics[width=0.4\textwidth]{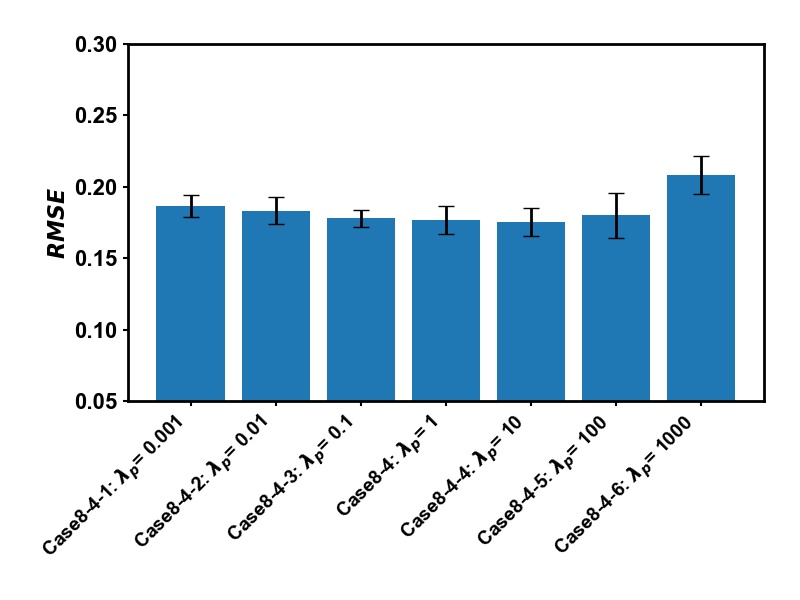}
    \label{figCa:Fig_CoefP_MRSE_mag} }
  \subfloat[$RMSE$ of Wind speed direction]{
    \includegraphics[width=0.4\textwidth]{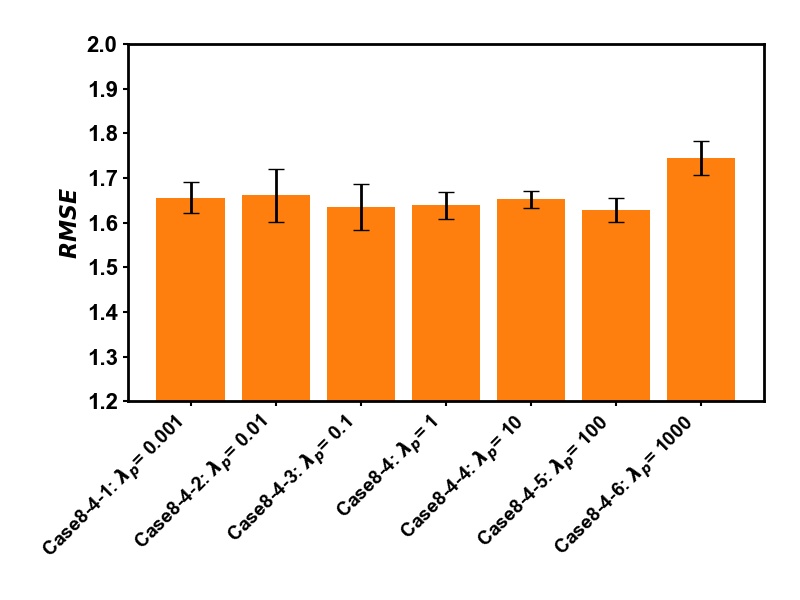} \label{figCb:Fig_CoefP_MRSE_dir} }
\caption{The $RMSE$ under various pressure weight $\lambda_{p}$. The $RMSE$ values represent the relative root mean square errors between the reconstructed wind flow fields generated by the trained PINN and the ground truth. These $RMSE$ values are computed as the averages of five independent training sessions, with error bars indicating the standard deviations across these sessions.}
\label{figC:CoefP_MRSE}
\end{figure*}

In Section \ref{sec:2.2_Assimilation}, the weights for all measurement data are considered equal, facilitating the explanation of the adjustment between data loss and equation loss weights.
However, in practical applications, the reliability of sensors might vary, or the interest in data from specific sensor regions may differ.
To accommodate this, the proposed framework includes an interface for finely tuning the weights assigned to each type of sensor data.
Specifically, Equation \ref{eq:loss_data} can be expressed as
\begin{equation}
\mathcal{L}_{data} = \lambda_{p}e_p + \lambda_{LoS}e_{LoS} + \lambda_{vec}e_{vec} + \lambda_{comp}e_{comp}
\label{eq:loss_data_sep}
\end{equation}
where $\lambda_{p}$, $\lambda_{LoS}$, $\lambda_{vec}$ and $\lambda_{comp}$ represent the weights for pressure data, LoS wind speeds, velocity vectors and velocity components, respectively.
This section demonstrates, using the fine-tuning of pressure data weights as an example, the impact of data weighting adjustments.
Building on Case 8-4, the weight for pressure data, $\lambda_{p}$ , is set to values of 0.001, 0.01, 0.1, 10, 100, and 1000, while the weights for velocity-related data, $\lambda_{LoS}$, $\lambda_{vec}$ and $\lambda_{comp}$, are all maintained at 1.
The remaining network training settings are kept consistent with those of Case 8-4, and five independent training sessions are conducted for each $\lambda_{p}$ value.
Figure \ref{figC:CoefP_MRSE} displays the $RMSE$ of the reconstructed flow fields by the trained PINN upon various $\lambda_{p}$.
The results align with the patterns observed in Section \ref{sec:3.4_Influence of coef}: moderate adjustments to the weights can enhance the accuracy of the reconstructed flow fields, while excessively increasing or decreasing the weights is detrimental.
Given that Case 8-4 already achieved a well-balanced data and equation framework, changes in the pressure data weight, $\lambda_{p}$, has limited impact on the accuracy of the reconstructed flow fields.
This appendix serves merely as an example to demonstrate the feasibility of fine-tuning data weights within the proposed framework.

\clearpage
\bibliographystyle{unsrt}  
\bibliography{references}

\end{document}